\documentclass[12pt]{iopart}

\usepackage{graphicx}

\begin{document}
\newcommand{\C}[1]{\mathcal{#1}}

\title{Analytical solutions of bound timelike geodesic orbits in Kerr spacetime}

\author{Ryuichi Fujita$^{1,2}$ and Wataru Hikida$^2$}

\address{${}^1$Theoretical Physics, Raman Research Institute, Bangalore 560 080, India}
\address{${}^2$Department of Earth and Space Science, Graduate School of
Science, Osaka University, Toyonaka, Osaka 560-0043, Japan
}
\ead{draone@rri.res.in}
\begin{abstract}
We derive the analytical solutions of the bound timelike geodesic orbits 
in Kerr spacetime. The analytical solutions are expressed 
in terms of the elliptic integrals using Mino time $\lambda$ 
as the independent variable. 
Mino time decouples the radial and polar motion of a particle
and hence leads to forms more useful to estimate 
three fundamental frequencies, 
radial, polar and azimuthal motion, 
for the bound timelike geodesics in Kerr spacetime. 
This paper gives the first derivation of 
the analytical expressions of the fundamental frequencies. 
This paper also gives the first derivation of 
the analytical expressions of all coordinates 
for the bound timelike geodesics using Mino time. 
These analytical expressions 
should be useful not only to investigate physical properties of Kerr geodesics
but more importantly to applications related to the estimation of 
gravitational waves from the extreme mass ratio inspirals.
\end{abstract}

\maketitle

\section{Introduction}
\label{sec:intro}
The Kerr black hole has been well studied 
since the discovery of the Kerr solution. 
It is an important topic not only in mathematical problems of
 general theory of relativity, but also for applications in 
astrophysics. 
Currently, there are many candidates for black holes 
in the universe and they have a wide range of mass scales 
ranging from  stellar mass scales to galactic nuclei mass scales~\cite{Narayan}. 

One of the  ways to investigate the properties 
of a  Kerr black hole spacetime is to study  geodesic motion in this 
background. 
Detailed works on the geodesic motion in black hole spacetimes 
are summarized in Chandrasekhar~\cite{Chandra}. 
In the weak field regime, at large distances from the black hole, 
the orbits of a particle are almost the same as that in Newtonian gravity. 
In the strong field regime, however, 
the orbits become more complicated and 
it is difficult to compare the orbits with that in Newtonian gravity. 
For the case of bound geodesics, 
this can be explained by mismatches between the
fundamental frequencies of radial, $\Omega_r$, polar, $\Omega_\theta$ and 
azimuthal-motion, $\Omega_\phi$. 
For example, $\Omega_\phi-\Omega_\theta$ shows the precession of the
 orbital plane and 
$\Omega_\phi-\Omega_r$ shows the precession of the orbital ellipse. 
Differences between the fundamental frequencies 
become larger as the particle goes into the strong gravity region 
around black hole horizon or separatrix, 
which is the boundary between stable and unstable orbits. 
These relativistic effects have been studied for some cases 
and some examples of extreme phenomena are found as follows. 

Wilkins~\cite{wilkins} derived the analytical expressions for the ratio
of the azimuthal frequency and the polar frequency,
 $\Omega_\phi/\Omega_\theta$, 
when a particle moves on both circular and non-equatorial orbits 
around the extreme Kerr black hole. 
He then  showed that the ratio becomes larger 
as the particle approaches the horizon
and found that the particle traces out a helix-like orbit on a sphere around
the black hole. 
He also pointed out that there exist horizon-skimming orbits which have 
the same radius as the horizon. 
Horizon-skimming orbits are also studied by numerical calculations 
including the effects of the emission of gravitational waves from a particle 
for circular and non-equatorial orbits~\cite{Hughes2} and 
for generic orbits~\cite{Barausse} around near-extremal Kerr black holes. 
Glampedakis and Kennefick~\cite{GK} numerically investigated the ratio
of the azimuthal frequency and the radial frequency, $\Omega_\phi/\Omega_r$, 
when a particle moves  both on eccentric and equatorial orbits around
the Kerr black hole. 
They found that the ratio becomes larger as the particle approaches the
 separatrix  and  the particle traces out a quasi-circular orbit around
 the periapsis 
before going back to the apoapsis. These orbits are called zoom-whirl orbits.

The above results show that the fundamental frequencies play an important role 
in  understanding bound geodesic orbits. 
However, the coupling of the $r$ and $\theta$ motions in the geodesic equation 
has prevented one from deriving the fundamental frequencies, 
$\Omega_r$, $\Omega_\theta$ and $\Omega_\phi$, 
for general bound geodesic orbits until recently. 
Using the elegant Hamilton-Jacobi formalism, 
Schmidt~\cite{Celestial} derived the fundamental frequencies 
without discussing the coupling of the $r$ and $\theta$ motions. 
Although his results show that we can expand an arbitrary function 
of the  particle's orbit  in a Fourier series, we can not estimate the 
Fourier components 
because of the coupling of the $r$ and $\theta$-motion. 
Mino~\cite{Mino:2003yg} showed that we can separate $r$ and $\theta$-motion 
if we use new time parameter $\lambda$ and 
derived the integral forms of the periods 
of both $r$ and $\theta$-motion with respect to $\lambda$, 
which is called Mino time. 
Combining Schmidt's method with Mino time, 
Drasco and Hughes~\cite{Drasco:2004} derived the fundamental
frequencies and showed how the Fourier components of arbitrary functions of 
orbits with respect to Mino time can be computed because of the decoupling of 
both $r$ and $\theta$ motions. They also showed how from these results 
using Mino time, the Fourier components with respect to coordinate time 
can also be derived. 
Thanks to these results, 
one can compute gravitational waves from binary systems in which 
a stellar mass compact star is moving on a general bound geodesic orbit 
around a supermassive black hole, the so-called 
extreme mass ratio inspirals(EMRIs)~\cite{Kostas_Review}. 
Gravitational waves from EMRIs are one of the main targets for space-based 
Laser Interferometer Space Antenna (LISA)~\cite{LISA}. 

In this paper, we derive analytical expressions for bound timelike geodesic 
orbits in Kerr spacetime using Mino time as the independent variable. 
Despite a lot of works on geodesic motion~\cite{Chandra}, 
the analytical expressions of null or timelike geodesics 
in Kerr spacetime are still important subjects. 
Fast and accurate computation of null geodesics in Kerr spacetime 
is required to study radiation which pass near black holes in 
accretion systems such as active galactic nuclei and X-ray binaries 
(see, for example, \cite{Rauch,Dexter} and references therein). 
Fast and accurate computation of timelike geodesics 
is also required to study gravitational waves from EMRIs and 
construct efficient templates for LISA data analysis. 
Rauch and Blandford gave tables which reduce 
the some integral forms of null geodesics 
to Legendre elliptic integrals~\cite{handbook} 
using the radial coordinate as the independent variable~\cite{Rauch}. 
They did not give the complete tables which reduce 
all the integral forms to the elliptic integrals 
because it was easier and faster to compute 
both $t$ and $\phi$ coordinates numerically 
when they studied the optical structure of the primary caustic around 
Kerr black hole. 
Using Carlson elliptic integrals~\cite{carlson1988} 
to calculate all coordinates of null geodesics, 
however, Dexter and Agol showed that they can compute null geodesics 
more efficiently than numerical integration method~\cite{Dexter}. 
Although they did not show analytical expressions of all coordinates 
of null geodesics since there are so many cases to be considered, they opened 
their numerical code to compute null geodesics semi-analytically 
in Kerr spacetime. 
In this paper, we show that we can easily derive the analytical 
expressions of bound timelike geodesics 
in terms of Legendre elliptic integrals 
if we properly transform the $r$ and $\theta$ variables. 
This is the first time that the analytical expressions of 
fundamental frequencies are derived. 
This is also the first time that the analytical expressions of 
all geodesic coordinates are derived using Mino time 
as the independent variable. 
These analytical expressions of bound timelike geodesic orbits 
with respect to Mino time are simpler than that in \cite{Dexter} 
for null geodesics and 
should be useful to investigate gravitational waves from EMRIs. 
The analytical solutions should also be helpful for investigations 
of bound geodesics in Kerr spacetime. 

This paper is organized as follows. 
In \sref{sec:geodesics}, we review Kerr geodesics using observer time. 
We then discuss Kerr geodesics in Mino time and derive the 
analytical expressions of 
the fundamental frequencies of bound geodesics in \sref{sec:Omega}. 
In \sref{sec:ana_orbit}, we derive the analytical expressions for 
bound geodesic 
orbits. We conclude with a brief summary in \sref{sec:summary}. 
In the Appendices, we discuss technical details of the  implementation
required to obtain the results in this paper. 
Throughout this paper, we use units with $G=c=1$.
\section{Geodesic Orbits in Kerr Spacetime}
\label{sec:geodesics}
The geodesic equations that describe a particle's orbits in Kerr spacetime are given by 
\begin{eqnarray}
\Sigma^2\left(\frac{\rmd r}{\rmd\tau}\right)^2 = R(r),\cr
\Sigma^2\left(\frac{\rmd\cos\theta}{\rmd\tau}\right)^2 = \Theta(\cos\theta),\cr
\Sigma\frac{\rmd t}{\rmd\tau} = T_{\rm r}(r)+T_\theta(\cos\theta) + a\C{L}_{z},\cr
\Sigma\frac{\rmd\phi}{\rmd\tau} = \Phi_{\rm r}(r)+\Phi_\theta(\cos\theta) - a\C{E}.
\label{eq:geodesic_tau}
\end{eqnarray}

The functions $R(r)$, $\Theta(\cos\theta)$, 
$T_{\rm r}(r)$, $T_{\theta}(\cos\theta)$,
$\Phi_{\rm r}(r)$ and $\Phi_{\theta}(\cos\theta)$ are defined by 
\begin{eqnarray*}
R(r)=[P(r)]^2-\Delta[r^2+(a\C{E}-\C{L}_{z})^2+\C{C}], \cr
\Theta(\cos\theta)=
\C{C} - (\C{C}+a^2(1-\C{E}^2)+\C{L}_{z}^2)\cos^2\theta
+ a^2(1-\C{E}^2)\cos^4\theta,\cr
T_{r}(r)=\frac{r^2+a^2}{\Delta}P(r),\qquad
T_{\theta}(\cos\theta) = -a^2\C{E}(1-\cos^2\theta),\cr
\Phi_{r}(r)=\frac{a}{\Delta}P(r),\qquad
\Phi_{\theta}(\cos\theta) =\frac{\C{L}_{z}}{1-\cos^2\theta},
\end{eqnarray*}
with $P(r)=\C{E}(r^2+a^2)-a\C{L}_{z}$, $\Sigma=r^2+a^2\cos^2\theta$ and 
$\Delta=r^2-2Mr+a^2$. Here $M$ and $a$ are the mass and 
the angular momentum of the black hole, respectively.
There are three constants of motion, 
$\C{E}$, $\C{L}_{z}$ and $\C{C}$, which  
are the energy, the z-component of the angular 
momentum and the Carter constant per unit mass, respectively. 
Using reasonable initial conditions  for the particle's orbit, 
we can derive the orbits using the proper time of the particle, $\tau$, 
by numerical integration. 
Dividing $\rmd r/\rmd\tau$, $\rmd\cos\theta/\rmd\tau$ and 
$\rmd\phi/\rmd\tau$ by $\rmd t/\rmd\tau$, 
we can also derive the orbits with coordinate time $t$ by numerical integration. 
When the orbits are bound to black hole, however, 
we have to take care of the turning points in the radial and the polar 
motion where the signs of $\rmd r/\rmd\tau$ and $\rmd\cos\theta/\rmd\tau$ change. 
These turning points correspond to periapsis and apoapsis for the radial motion, 
and $\theta_{\rm min}$ and $\pi-\theta_{\rm min}$ for the polar motion, 
where $\theta_{\rm min}$ is the minimum value of $\theta$.
We need smaller stepsizes to resolve the derivatives around turning points. 
We can avoid this problem by introducing new variables for
 the radial and polar motion, 
$r=pM/(1+e\cos\psi)$ and $\cos\theta=\cos\theta_{\rm min}\cos\chi$, 
where $p$ is semilatus rectum and $e$ is eccentricity~\cite{Drasco:2004}. 
Using these new variables, $\psi$ and $\chi$, we can estimate the orbits
accurately.

There exists three fundamental frequencies, 
$\Omega_r$, $\Omega_\theta$ and $\Omega_\phi$, for bound Kerr geodesics. 
However, it is difficult to estimate the fundamental frequencies 
using (\ref{eq:geodesic_tau}) because of the coupling of the
 $r$ and $\theta$-motions. 
For instance, we immediately face a difficulty when we estimate
 $\Omega_r$ using $\rmd r/\rmd t=(\rmd r/\rmd\tau)(\rmd t/\rmd\tau)^{-1}$
 because $r$ and $\theta$ asynchronously pass their 
turning points. 
\section{Fundamental Frequencies of bound geodesics}
\label{sec:Omega}
We now proceed to derive the analytical expressions for 
the fundamental frequencies of bound geodesic orbits using Mino time. 
In \sref{sec:mino}, we briefly describe the Kerr geodesics in Mino time 
and then show how to derive the analytical expressions for
the fundamental frequencies in \sref{sec:omega_r_theta}
and \sref{sec:omega_t_phi}. 
In \sref{sec:check_omega}, we will check the analytical expressions 
by comparing them with earlier literature. 
\subsection{Geodesics in Mino time}
\label{sec:mino}
Using Mino time, $\lambda=\int \rmd\tau/\Sigma$, 
the geodesic equations become
\begin{eqnarray}
\left(\frac{\rmd r}{\rmd\lambda}\right)^2 = R(r),\cr
\left(\frac{\rmd\cos\theta}{\rmd\lambda}\right)^2 = \Theta(\cos\theta),\cr
\frac{\rmd t}{\rmd\lambda} = T_{\rm r}(r)+T_\theta(\cos\theta) + a\C{L}_{z},\cr
\frac{\rmd\phi}{\rmd\lambda} = \Phi_{\rm r}(r)+\Phi_\theta(\cos\theta) - a\C{E}.
\label{eq:geodesic_lam}
\end{eqnarray}

It should be noted that, in (\ref{eq:geodesic_lam}), 
$\rmd r/\rmd\lambda$ depends only on $r$ and 
$\rmd\cos\theta/\rmd\lambda$ depends 
only on $\cos\theta$. 
Thus the equations for radial and polar motion are decoupled. 
For the bound orbits, $r(\lambda)$ and $\cos\theta(\lambda)$ become
periodic functions which are independent of each other. 
The fundamental periods for the radial and polar motion,
$\Lambda_r$ and $\Lambda_\theta$, with respect to $\lambda$ are given by 
\begin{eqnarray}
\Lambda_r=2\int_{\rm r_{\rm min}}^{\rm r_{\rm max}}\frac{\rm d r}{\sqrt{R(r)}},
\qquad
\Lambda_\theta=4\int_{0}^{\cos\theta_{\rm min}}
\frac{\rmd\cos\theta}{\sqrt{\Theta(\cos\theta)}},
\label{eq:lam_r_theta}
\end{eqnarray}
where 
\begin{eqnarray}
r_{\rm min} = \frac{pM}{1+{e}},\qquad
r_{\rm max} = \frac{pM}{1-{e}},\qquad
\theta_{\rm inc} + ({\rm sgn}\, \C{L}_{z})\, \theta_{\rm min} = \frac{\pi}{2}.
\label{eq:zero_r_theta}
\end{eqnarray}
Here $r_{\rm min}$ and $r_{\rm max}$ are the periapsis and apoapsis for the radial motion 
respectively, 
and $\theta_{\rm inc}$  the inclination angle from the equatorial plane of black hole.
Of course, $(\C{E},\C{L}_{z},\C{C})$ are described by these 
orbital parameters $(p,{e},\theta_{\rm inc})$ and 
given in \cite{Celestial,Drasco:2004}.
The angular frequencies of the radial and the polar motion then become
\begin{eqnarray}
\Upsilon_r=\frac{2\pi}{\Lambda_r},\qquad
\Upsilon_\theta=\frac{2\pi}{\Lambda_\theta}.
\end{eqnarray}

We also note that both 
$\rmd t/\rmd\lambda$ and $\rmd\phi/\rmd\lambda$ in (\ref{eq:geodesic_lam})
are the sum of a function of $r$ and a function of $\cos\theta$.
Then each equations are integrated as
\begin{eqnarray}
t(\lambda) = \Gamma\lambda + t^{(r)}(\lambda)+ t^{(\theta)}(\lambda),\qquad
\phi(\lambda) = \Upsilon_{\phi} \lambda+
\phi^{(r)}(\lambda)+ \phi^{(\theta)}(\lambda),
\label{eq:t_phi_decom}
\end{eqnarray}
where $\Gamma$ and $\Upsilon_{\phi}$ are 
the frequencies of coordinate time $t$ and $\phi$ 
with respect to $\lambda$ respectively, which are given by
\begin{eqnarray}
\Gamma = \Upsilon_{t^{(r)}}+\Upsilon_{t^{(\theta)}}+a\C{L}_{z},\qquad
&\Upsilon_{\phi} &= 
\Upsilon_{\phi^{(r)}}+\Upsilon_{\phi^{(\theta)}} - a\C{E},\cr
\Upsilon_{t^{(r)}} = \left<T_r(r)\right>_{\lambda},\qquad
&\Upsilon_{t^{(\theta)}} &= \left<T_\theta(\cos\theta)\right>_{\lambda},\cr
\Upsilon_{\phi^{(r)}} = \left<\Phi_r(r)\right>_{\lambda},\qquad
&\Upsilon_{\phi^{(\theta)}} &=
\left<\Phi_\theta(\cos\theta)\right>_{\lambda},
\label{eq:upsilon_t_phi}
\end{eqnarray}
where $\left<\cdots\right>_{\lambda}\equiv \lim_{\Delta\lambda\rightarrow\infty}(2\Delta\lambda)^{-1}\int_{-\Delta\lambda}^{\Delta\lambda}d\lambda\cdots$ 
represents infinite time average
with respect to $\lambda$,
and $t^{(r)/(\theta)}$ and $\phi^{(r)/(\theta)}$ satisfy
\begin{eqnarray}
\frac{\rmd t^{(r)}}{\rmd\lambda} = T_{r}(r) - \Upsilon_{t^{(r)}},\qquad
\frac{\rmd t^{(\theta)}}{\rmd\lambda} = T_{\theta}(\cos\theta) -\Upsilon_{t^{(\theta)}},\cr
\frac{\rmd\phi^{(r)}}{\rmd\lambda} = \Phi_{r}(r)-\Upsilon_{\phi^{(r)}},\qquad
\frac{\rmd\phi^{(\theta)}}{\rmd\lambda} = \Phi_{\theta}(\cos\theta)-\Upsilon_{\phi^{(\theta)}}.
\label{eq:dt_dphi_dlam}
\end{eqnarray}

Equation (\ref{eq:t_phi_decom}) 
shows that both $t(\lambda)$ and $\phi(\lambda)$ 
consist of two distinct parts. 
The first term represents an accumulation over $\lambda$-time and 
the last two terms represent oscillations around it with periods  
$2\pi/\Upsilon_{r}$ and $2\pi/\Upsilon_{\theta}$. 
We note that the frequencies with respect to $\lambda$ are related to 
the frequencies with distant observer time as~\cite{Drasco:2004}
\begin{eqnarray}
\Omega_{r}=\frac{\Upsilon_{r}}{\Gamma},\qquad
\Omega_{\theta}=\frac{\Upsilon_{\theta}}{\Gamma},\qquad
\Omega_{\phi}=\frac{\Upsilon_{\phi}}{\Gamma}.
\label{eq:Omega_r_th_phi}
\end{eqnarray}

In the following subsections, 
\sref{sec:omega_r_theta} and 
\sref{sec:omega_t_phi}, 
we discuss the analytical expressions for these frequencies. 
And we discuss the analytical expressions of the orbits, 
$r(\lambda)$, $\cos\theta(\lambda)$, $t(\lambda)$ and $\phi(\lambda)$, 
in \sref{sec:ana_orbit}.
\subsection{Frequencies of $r$ and $\theta$-motion}
\label{sec:omega_r_theta}
In this subsection, we derive the analytical expressions for 
the frequencies of $r$ and $\theta$-motion, $\Upsilon_r$ and $\Upsilon_\theta$, 
using (\ref{eq:lam_r_theta}). 
As explained in \sref{sec:geodesics}, 
$R(r)$ and $\Theta(\cos\theta)$ become zero when $r$ and $\cos\theta$ go 
through the turning points, 
$r_{\rm min}$, $r_{\rm max}$ and $\pm\cos\theta_{\rm min}$, respectively.
Thus we usually transform the variables, 
$r$ and $\cos\theta$, to avoid divergences in the numerical calculation. 
However we know that (\ref{eq:lam_r_theta}) can be expressed in terms of 
the elliptic integrals since 
both $R(r)$ and $\Theta(\cos\theta)$ are fourth order polynomials~\cite{Bardeen}.
It is useful if we know the four zero points of both 
$R(r)$ and $\Theta(\cos\theta)$ to express (\ref{eq:lam_r_theta}) 
in terms of the elliptic integrals. 
We rewrite $R(r)$ and $\Theta(\cos\theta)$ as~\cite{Drasco:2004}
\begin{eqnarray}
R(r)=(1 - \C{E}^2)(r_1 - r)(r - r_2)(r - r_3)(r - r_4), \cr
\Theta(\cos\theta)=\C{L}_{z}^2\epsilon_{0}(z_{-}-\cos^2\theta)(z_{+}-\cos^2\theta),
\label{eq:factor_R_Theta}
\end{eqnarray}
where 
\begin{eqnarray}
\fl
 r_{1} = \frac{pM}{1-{e}},\qquad
 r_{2} &= \frac{pM}{1+{e}},\qquad
 r_{3} = \frac{(A+B)+\sqrt{(A+B)^2-4AB}}{2},\qquad
 r_{4} = \frac{AB}{r_3},\cr
 A+B &= \frac{2M}{1-{\C{E}}^2} - (r_1+r_2),\qquad
 AB  = \frac{a^2\C{C}}{(1-{\C{E}}^2)\,r_1r_2},
\end{eqnarray}
and where $\epsilon_{0}=a^2(1-\C{E}^2)/\C{L}_{z}^2$, 
$z_{-}=\cos^2\theta_{\rm min}$ and $z_{+}=\C{C}/(\C{L}_{z}^2\epsilon_0z_{-})$.
We note that 
two zero points,
$r_{1}$ and $r_{2}$, of $R(r)$ are apoapsis and periapis respectively 
and two zero points, $z_{-}$ and $-z_{-}$, of $\Theta(\cos\theta)$ 
are $\theta_{\rm min}$ and $\pi-\theta_{\rm min}$ respectively. 
These zero points correspond to turning points, 
defined in (\ref{eq:zero_r_theta}), of radial and polar motion. 
But the other two zero points of both $R(r)$ and $\Theta(\cos\theta)$, 
$r_{3}$, $r_{4}$ and $\pm z_{+}$, do not correspond to turning points of 
radial and polar motion.

Using (\ref{eq:factor_R_Theta}), 
we can express (\ref{eq:lam_r_theta}) in terms of the elliptic integrals as
\begin{eqnarray}
\int^r_{r_2} \frac{\rmd r'}{\sqrt{R(r')}}=
\frac{2}{\sqrt{(1-{\C{E}}^2)(r_1-r_3)(r_2-r_4)}}F(\arcsin y_r,k_r),\cr
\int^{\cos\theta}_0\frac{\rmd\cos\theta '}{\sqrt{\Theta(\cos\theta ')}}=
\frac{1}{\C{L}_{z}\sqrt{\epsilon_0 z_{+}}}F(\arcsin y_\theta,k_\theta),
\label{eq:ellipt_r_th}
\end{eqnarray}
where 
\begin{eqnarray}
y_r=\sqrt{\frac{r_1-r_3}{r_1-r_2}\frac{r-r_2}{r-r_3}},\qquad
k_{r} = \sqrt{\frac{r_1-r_2}{r_1-r_3}\frac{r_3-r_4}{r_2-r_4}},\cr
y_\theta=\frac{\cos\theta}{\sqrt{z_{-}}},\qquad
k_{\theta} =\sqrt{\frac{z_{-}}{z_{+}}},
\label{eq:yr_yth}
\end{eqnarray}
and $F(\varphi,k)$ is the incomplete elliptic integral of the first kind defined by 
\begin{eqnarray}
F(\varphi,k)=\int_0^\varphi\frac{\rmd\theta}{\sqrt{1-k^2\sin^2\theta}}
=\int_0^{\sin\varphi}\frac{\rmd y}{\sqrt{(1-y^2)(1-k^2y^2)}}. 
\end{eqnarray}
In the following, we describe both the elliptic integrals and the elliptic functions
using the notation in \cite{Recipes}.
The orbital frequencies of radial and polar motion 
with respect to $\lambda$ are then given by
\begin{eqnarray}
\Upsilon_{r} =
\frac{\pi\sqrt{(1-{\C{E}}^2)(r_1-r_3)(r_2-r_4)}}{2K(k_r)},\qquad
\Upsilon_{\theta} =
\frac{\pi\C{L}_{z}\sqrt{\epsilon_0 z_{+}}}{2K(k_{\theta})}.
\label{eq:Up_r_th}
\end{eqnarray}
Here $K(k)$ is the complete elliptic integral of the first kind
defined by $K(k)=F(\pi/2,k)$. We note that the analysis of $\theta$-motion here 
is similar to that of Drasco and Hughes~\cite{Drasco:2004}. 
Though they  used a different transformation of $\cos\theta$,
our  expression for  $\Upsilon_\theta$ in this subsection agrees with
their final result. 
\subsection{Frequencies of $t$ and $\phi$-motion}
\label{sec:omega_t_phi}
In this subsection, we derive the analytical expressions for 
the frequencies of $t$ and $\phi$-motion, $\Gamma$ and $\Upsilon_\phi$, 
using (\ref{eq:upsilon_t_phi}). 
Since the  $r$ and $\theta$-motion decouple in Mino time, we can rewrite 
the infinite time average in (\ref{eq:upsilon_t_phi})  as an
average over an orbital period, $\Lambda_r$ or $\Lambda_\theta$, as
\begin{eqnarray}
\Upsilon_{t^{(r)}} &= \frac{2}{\Lambda_r}\int^{r_1}_{r_2} \frac{T_r(r)}{\sqrt{R(r)}}\rmd r,\qquad
\Upsilon_{t^{(\theta)}} = \frac{4}{\Lambda_\theta}\int^{\sqrt{z_{-}}}_0\frac{T_\theta(\cos\theta)}{\sqrt{\Theta(\cos\theta)}}\rmd\cos\theta,\cr
\Upsilon_{\phi^{(r)}} &= \frac{2}{\Lambda_r}\int^{r_1}_{r_2} \frac{\Phi_r(r)}{\sqrt{R(r)}}\rmd r,\qquad
\Upsilon_{\phi^{(\theta)}} = \frac{4}{\Lambda_\theta}\int^{\sqrt{z_{-}}}_0\frac{\Phi_\theta(\cos\theta)}{\sqrt{\Theta(\cos\theta)}}\rmd\cos\theta.
\end{eqnarray}
It is straightforward to express $\Upsilon_{t^{(\theta)}}$ and 
$\Upsilon_{\phi^{(\theta)}}$ in terms of the elliptic integrals 
if we use $y_\theta$ in (\ref{eq:yr_yth}).
\begin{eqnarray}
\Upsilon_{t^{(\theta)}} &= -\frac{2a^2\C{E}\Upsilon_{\theta}}{\pi\C{L}_{z}\sqrt{\epsilon_0 z_{+}}}
\left[(1-z_{+})K(k_\theta)+z_{+}E(\frac{\pi}{2},k_\theta)\right],\cr
\Upsilon_{\phi^{(\theta)}} &= \frac{2\Upsilon_\theta}{\pi\sqrt{\epsilon_0 z_{+}}}\Pi(\frac{\pi}{2},-z_{-},k_\theta),
\end{eqnarray}
where $E(\varphi,k)$ is the incomplete elliptic integral of the second kind and 
$\Pi(\varphi,c,k)$ is the incomplete elliptic integral of the third kind defined by 
\begin{eqnarray}
E(\varphi,k)&=\int_0^\varphi\sqrt{1-k^2\sin^2\theta}\rmd\theta,\cr
\Pi(\varphi,c,k)&=
\int_0^\varphi\frac{\rmd\theta}{(1+c\sin^2\theta)\sqrt{1-k^2\sin^2\theta}}.
\end{eqnarray}
Note that, $E(\pi/2,k)$ is the complete elliptic integral of the second kind
 and 
$\Pi(\pi/2,c,k)$ is the complete elliptic integral of the third kind. 
In the followings, we describe $E(\pi/2,k)$ as $E(k)$, and 
$\Pi(\pi/2,c,k)$ as $\Pi(c,k)$.

On the other hand, we have to rewrite $T_{r}(r)$ and $\Phi_{r}(r)$ 
in order to express $\Upsilon_{t^{(r)}}$ and $\Upsilon_{\phi^{(r)}}$ 
in terms of the elliptic integrals. 
Performing partial fraction decomposition, 
we decompose $T_{r}(r)$ and $\Phi_{r}(r)$ as follows.
\begin{eqnarray}
T_{r}(r)&=&
\C{E}r^2+2M\C{E}r+\frac{2M}{r_{+}-r_{-}}\left\{\frac{(4M^2\C{E}-a\C{L}_{z})r_{+}-2Ma^2\C{E}}{r-r_{+}}-(+\leftrightarrow -)\right\}\cr
&&+(a^2+4M^2)\C{E}-a\C{L}_{z}\qquad (|a|\neq M),\cr
&=&\C{E}r^2+2M\C{E}r+\frac{2M(4M^2\C{E}-a\C{L}_{z})}{r-M}+\frac{2M^2(2M^2\C{E}-a\C{L}_{z})}{(r-M)^2}\cr
&&+(a^2+4M^2)\C{E}-a\C{L}_{z}\qquad (|a|= M),\cr
\Phi_{r}(r)&=&
\frac{a}{r_{+}-r_{-}}\left\{\frac{2M\C{E}r_{+}-a\C{L}_{z}}{r-r_{+}}-(+\leftrightarrow -)\right\}+a\C{E}\qquad (|a|\neq M),\cr
&=&\frac{2Ma\C{E}}{r-M}+\frac{a(2M^2\C{E}-a\C{L}_{z})}{(r-M)^2}+a\C{E}\qquad (|a|= M),
\label{eq:tr_phir_decom}
\end{eqnarray}
with $r_\pm=M\pm\sqrt{M^2-a^2}$. 
From (\ref{eq:yr_yth}), we find $r=r_3+(r_2-r_3)/(1-h_{r}y_{r}^2)$, where
$h_r=(r_1-r_2)/(r_1-r_3)$. Then it is straightforward to compute 
$\int_{r_2}^{r}r'\rmd r'/\sqrt{R(r')}$, 
which is composed in $\Upsilon_{t^{(r)}}$, as
\begin{eqnarray}
\int^r_{r_2} \frac{r'}{\sqrt{R(r')}}\rmd r' =
\frac{2\left[r_3F(\arcsin y_r,k_r)+(r_2-r_3)\Pi(\arcsin y_r,-h_r,k_r)\right]}{\sqrt{(1-{\C{E}}^2)(r_1-r_3)(r_2-r_4)}}.
\end{eqnarray}
We show the results of the other terms of 
$\int_{r_2}^{r}T_{r}(r')\rmd r'/\sqrt{R(r')}$ and 
$\int_{r_2}^{r}\Phi_{r}(r')\rmd r'/\sqrt{R(r')}$ in \ref{sec:integral_formula}. 
Using results quoted there, 
we derive $\Gamma$ and $\Upsilon_\phi$ as\footnote{As pointed out 
in~\cite{SY2011}, in the first version of this article there was a typo 
in $\Gamma$ : the square closing bracket 
in the third line should be moved to include the next term, 
i.e. $(r_1-r_3)(r_2-r_4)E(k_r)$. We thank C. F. Sopuerta and N. Yunes 
for pointing out the typo.}
\begin{eqnarray}
\fl
\Gamma = 4M^2\C{E}+\frac{2a^2\C{E}z_{+}\Upsilon_{\theta}}{\pi\C{L}_{z}\sqrt{\epsilon_0 z_{+}}}
\left[K(k_\theta)-E(k_\theta)\right]
+\frac{2\Upsilon_{r}}{\pi\sqrt{(1-\C{E}^2)(r_1-r_3)(r_2-r_4)}}\cr
\times\left\{\frac{\C{E}}{2}\left[(r_3(r_1+r_2+r_3)-r_1r_2)K(k_r)\right.\right.\cr
+(r_2-r_3)(r_1+r_2+r_3+r_4)\Pi(-h_r,k_r)\cr
+\left.(r_1-r_3)(r_2-r_4)E(k_r)\right]+2M\C{E}\left[r_3K(k_r)+(r_2-r_3)\Pi(-h_r,k_r)\right]\cr
+\left.\frac{2M}{r_{+}-r_{-}}\left[\frac{(4M^2\C{E}-a\C{L}_{z})r_{+}-2Ma^2\C{E}}{r_3-r_{+}}
\left(K(k_r)-\frac{r_2-r_3}{r_2-r_{+}}\Pi(-h_{+},k_r)
\right)\right.\right.\cr
-\left.\left.(+\leftrightarrow -)\right]\right\},\cr
\fl
\Upsilon_{\phi} = \frac{2\Upsilon_{\theta}}{\pi\sqrt{\epsilon_0 z_{+}}}
\Pi(-z_{-},k_\theta)+\frac{2a\Upsilon_{r}}{\pi(r_{+}-r_{-})\sqrt{(1-\C{E}^2)(r_1-r_3)(r_2-r_4)}}\cr
\times\left\{\frac{2M\C{E}r_{+}-a\C{L}_{z}}{r_3-r_{+}}\left[K(k_r)-\frac{r_2-r_3}{r_2-r_{+}}\Pi(-h_{+},k_r)\right]-(+\leftrightarrow -)\right\},
\label{eq:Up_t_phi}
\end{eqnarray}
where $h_\pm=(r_1-r_2)(r_3-r_\pm)/[(r_1-r_3)(r_2-r_\pm)]$. 
Combining (\ref{eq:Omega_r_th_phi}), (\ref{eq:Up_r_th}) and (\ref{eq:Up_t_phi}), 
we can derive the orbital frequencies with respect to observer time, 
$\Omega_r$, $\Omega_\theta$ and $\Omega_\phi$. 
It should be noted that (\ref{eq:Up_t_phi}) is not valid for the case 
$|a|=M$ since there exists divergent terms in $1/(r_{+}-r_{-})$. 
We show the expressions of $\Gamma$ and $\Upsilon_\phi$ 
for the case $|a|=M$ in \ref{sec:a_M}. 
\subsection{Consistency check of the fundamental frequencies}
\label{sec:check_omega}
We can compare the expressions for the fundamental frequencies, 
$\Omega_r$, $\Omega_\theta$ and $\Omega_\phi$, in this section with 
that in earlier literature for some limiting cases. 
We can compare $\Omega_\phi$ 
for the case $e=0$ and $\theta_{\rm inc}=0$ with \cite{BPT}, 
$\Omega_\phi/\Omega_r$ 
in the case $a=0$, $e\neq 0$ and $\theta_{\rm inc}=0$ with \cite{CKP}, 
and both $\Omega_\phi$ and $\Omega_\theta$ 
in the case $e=0$ and $\theta_{\rm inc}\neq 0$ with \cite{Hughes1}.
It is a good check on our results that 
our more general analytical expressions for the fundamental frequencies 
are consistent with earlier work for the limiting cases. 

We can check our results for more general cases. 
In \tref{tab:PN1} and \tref{tab:PN2}, 
we compare our results with \cite{Ganz} in which 
the analytical expressions of the orbital frequencies are derived 
in terms of both post-Newtonian and small eccentricity expansions 
up through $O(v^{5},\,{e}^2)$, where $v^2=M/p$. 
In \tref{tab:PN1} and \tref{tab:PN2}, 
we check our results for the cases $a\neq M$ and $a=M$ respectively. 
We find that relative errors are always less than $10^{-4}$ 
when we compare the results for the cases $p=100M$ and 
$0.01\le e\le 0.09$. 
Since the fundamental frequencies in \cite{Ganz} are derived up through 
$O(v^{5},\,{e}^2)$, these relative errors, less than $10^{-4}$, 
show the consistency of our results with \cite{Ganz}. 
In \tref{tab:Numerical}, we compare our results with numerical 
integration method for the eccentric and inclined orbits such that 
$p=6M$, $e=0.7$, $\theta_{\rm inc}=20^{\circ}$ and $a=0.9M$ or $a=M$. 
In numerical integration method, 
we use the trapezium rule to compute the fundamental frequencies. 
We can compute very accurately if we use 
the trapezium rule for the numerical integration of a periodic function. 
Then we find that the analytical expressions of the fundamental 
frequencies in this section agree with the results of 
numerical integration method. The relative errors are less than 
$10^{-15}$ in double precision calculation. These facts show that 
the analytical expressions in this section are correct in the cases of 
generic bound orbits. 

\begin{table}[htbp]
\caption{
Comparison of the orbital frequencies, 
$\Omega_r$, $\Omega_\theta$ and $\Omega_\phi$, 
derived using analytical expressions in this work and the analytical 
post-Newtonian expressions for 
orbits which are slightly eccentric but greatly inclined~\cite{Ganz} 
in the case of $a=0.9M$ and $p=100M$.
Our results are consistent with post-Newtonian results. 
Relative errors of the orbital frequencies are always less than $10^{-4}$. 
}\label{tab:PN1}
\begin{center}
{\tiny
\begin{tabular}{c c c c c c c c}
\br
  $\ \ {e}\ \ $ & $\theta_{\rm inc}$
& $\Omega_r^{\rm This\,\,work} $
& $\Omega_\theta^{\rm This\,\,work} $
& $\Omega_\phi^{\rm This\,\,work} $
& $\Omega_r^{\rm Post-Newton} $
& $\Omega_\theta^{\rm Post-Newton} $
& $\Omega_\phi^{\rm Post-Newton} $\\
\mr
$0.01$ & $20^{\circ}$ &
$\ 9.71944\times 10^{-4}\ $&
$\ 9.97417\times 10^{-4}\ $& 
$\ 9.99102\times 10^{-4}\ $&
$\ 9.71979\times 10^{-4}\ $&
$\ 9.97416\times 10^{-4}\ $& 
$\ 9.99102\times 10^{-4}\ $\\
$0.01$ & $45^{\circ}$ &
$\ 9.71338\times 10^{-4}\ $&
$\ 9.97974\times 10^{-4}\ $& 
$\ 9.99688\times 10^{-4}\ $&
$\ 9.71367\times 10^{-4}\ $ &
$\ 9.97975\times 10^{-4}\ $ &
$\ 9.99689\times 10^{-4}\ $\\
$0.01$ & $70^{\circ}$ &
$\ 9.70356\times 10^{-4}\ $&
$\ 9.98889\times 10^{-4}\ $& 
$\ 1.00065\times 10^{-3}\ $&
$\ 9.70379\times 10^{-4}\ $ &
$\ 9.98891\times 10^{-4}\ $ & 
$\ 1.00065\times 10^{-3}\ $\\\mr
$0.05$ & $20^{\circ}$ &
$\ 9.68513\times 10^{-4}\ $&
$\ 9.93895\times 10^{-4}\ $& 
$\ 9.95575\times 10^{-4}\ $&
$\ 9.68558\times 10^{-4}\ $&
$\ 9.93905\times 10^{-4}\ $& 
$\ 9.95586\times 10^{-4}\ $\\
$0.05$ & $45^{\circ}$ &
$\ 9.67910\times 10^{-4}\ $&
$\ 9.94452\times 10^{-4}\ $& 
$\ 9.96160\times 10^{-4}\ $&
$\ 9.67949\times 10^{-4}\ $ &
$\ 9.94464\times 10^{-4}\ $ &
$\ 9.96173\times 10^{-4}\ $\\
$0.05$ & $70^{\circ}$ &
$\ 9.66934\times 10^{-4}\ $&
$\ 9.95366\times 10^{-4}\ $& 
$\ 9.97118\times 10^{-4}\ $&
$\ 9.66968\times 10^{-4}\ $ &
$\ 9.95380\times 10^{-4}\ $ & 
$\ 9.97132\times 10^{-4}\ $\\\mr
$0.09$ & $20^{\circ}$ &
$\ 9.60520\times 10^{-4}\ $&
$\ 9.85694\times 10^{-4}\ $& 
$\ 9.87360\times 10^{-4}\ $&
$\ 9.60669\times 10^{-4}\ $&
$\ 9.85810\times 10^{-4}\ $& 
$\ 9.87478\times 10^{-4}\ $\\
$0.09$ & $45^{\circ}$ &
$\ 9.59926\times 10^{-4}\ $&
$\ 9.86249\times 10^{-4}\ $& 
$\ 9.87943\times 10^{-4}\ $&
$\ 9.60069\times 10^{-4}\ $ &
$\ 9.86368\times 10^{-4}\ $ &
$\ 9.88063\times 10^{-4}\ $\\
$0.09$ & $70^{\circ}$ &
$\ 9.58963\times 10^{-4}\ $&
$\ 9.87162\times 10^{-4}\ $& 
$\ 9.88899\times 10^{-4}\ $&
$\ 9.59101\times 10^{-4}\ $ &
$\ 9.87282\times 10^{-4}\ $ & 
$\ 9.89020\times 10^{-4}\ $\\
\br
\end{tabular}
}
\end{center}
\end{table}

\begin{table}[htbp]
\caption{
Comparison of the orbital frequencies, 
$\Omega_r$, $\Omega_\theta$ and $\Omega_\phi$, 
derived using analytical expressions in this work and the analytical 
post-Newtonian expressions for 
orbits which are slightly eccentric but greatly inclined~\cite{Ganz} 
in the case of $a=M$ and $p=100M$.
Our results are consistent with post-Newtonian results. 
Relative errors of the orbital frequencies are always less than $10^{-4}$. 
}\label{tab:PN2}
\begin{center}
{\tiny
\begin{tabular}{c c c c c c c c}
\br
  $\ \ {e}\ \ $ & $\theta_{\rm inc}$
& $\Omega_r^{\rm This\,\,work} $
& $\Omega_\theta^{\rm This\,\,work} $
& $\Omega_\phi^{\rm This\,\,work} $
& $\Omega_r^{\rm Post-Newton} $
& $\Omega_\theta^{\rm Post-Newton} $
& $\Omega_\phi^{\rm Post-Newton} $\\
\mr
$0.01$ & $20^{\circ}$ &
$\ 9.72213\times 10^{-4}\ $&
$\ 9.97159\times 10^{-4}\ $& 
$\ 9.99017\times 10^{-4}\ $&
$\ 9.72250\times 10^{-4}\ $&
$\ 9.97157\times 10^{-4}\ $& 
$\ 9.99017\times 10^{-4}\ $\\
$0.01$ & $45^{\circ}$ &
$\ 9.71546\times 10^{-4}\ $&
$\ 9.97769\times 10^{-4}\ $& 
$\ 9.99663\times 10^{-4}\ $&
$\ 9.71575\times 10^{-4}\ $ &
$\ 9.97770\times 10^{-4}\ $ &
$\ 9.99664\times 10^{-4}\ $\\
$0.01$ & $70^{\circ}$ &
$\ 9.70462\times 10^{-4}\ $&
$\ 9.98777\times 10^{-4}\ $& 
$\ 1.00073\times 10^{-3}\ $&
$\ 9.70484\times 10^{-4}\ $ &
$\ 9.98780\times 10^{-4}\ $ & 
$\ 1.00073\times 10^{-3}\ $\\\mr
$0.05$ & $20^{\circ}$ &
$\ 9.68780\times 10^{-4}\ $&
$\ 9.93637\times 10^{-4}\ $& 
$\ 9.95490\times 10^{-4}\ $&
$\ 9.68827\times 10^{-4}\ $&
$\ 9.93647\times 10^{-4}\ $& 
$\ 9.95501\times 10^{-4}\ $\\
$0.05$ & $45^{\circ}$ &
$\ 9.68117\times 10^{-4}\ $&
$\ 9.94247\times 10^{-4}\ $& 
$\ 9.96134\times 10^{-4}\ $&
$\ 9.68157\times 10^{-4}\ $ &
$\ 9.94259\times 10^{-4}\ $ &
$\ 9.96147\times 10^{-4}\ $\\
$0.05$ & $70^{\circ}$ &
$\ 9.67039\times 10^{-4}\ $&
$\ 9.95254\times 10^{-4}\ $& 
$\ 9.97196\times 10^{-4}\ $&
$\ 9.67072\times 10^{-4}\ $ &
$\ 9.95268\times 10^{-4}\ $ & 
$\ 9.97210\times 10^{-4}\ $\\\mr
$0.09$ & $20^{\circ}$ &
$\ 9.60784\times 10^{-4}\ $&
$\ 9.85436\times 10^{-4}\ $& 
$\ 9.87273\times 10^{-4}\ $&
$\ 9.60935\times 10^{-4}\ $&
$\ 9.85553\times 10^{-4}\ $& 
$\ 9.87391\times 10^{-4}\ $\\
$0.09$ & $45^{\circ}$ &
$\ 9.60130\times 10^{-4}\ $&
$\ 9.86045\times 10^{-4}\ $& 
$\ 9.87916\times 10^{-4}\ $&
$\ 9.60274\times 10^{-4}\ $ &
$\ 9.86164\times 10^{-4}\ $ &
$\ 9.88036\times 10^{-4}\ $\\
$0.09$ & $70^{\circ}$ &
$\ 9.59067\times 10^{-4}\ $&
$\ 9.87050\times 10^{-4}\ $& 
$\ 9.88975\times 10^{-4}\ $&
$\ 9.59204\times 10^{-4}\ $ &
$\ 9.87171\times 10^{-4}\ $ & 
$\ 9.89097\times 10^{-4}\ $\\
\br
\end{tabular}
}
\end{center}
\end{table}

\begin{table}[htbp]
\caption{
Comparison of the orbital frequencies, 
$\Omega_r$, $\Omega_\theta$ and $\Omega_\phi$, 
derived using analytical expressions in this work and 
numerical integration method in the case of 
$p=6M$, $e=0.7$, $\theta_{\rm inc}=20^{\circ}$ and $a=0.9M$ or $a=M$.
Our results are consistent with numerical integration method.
Relative errors of the orbital frequencies agree with 
the accuracy of double precision calculation. 
}\label{tab:Numerical}
\begin{center}
{\scriptsize
\begin{tabular}{c c c c c}
\br
$a/M$&
$\Omega_{r,\theta,\phi}$& 
{\rm This work} & 
{\rm Numerical integration} &
{\rm Absolute value of relative error}\\
\mr
$0.9$&
$\Omega_r$&
$1.8928532285101992\times 10^{-2}$&
$1.8928532285101982\times 10^{-2}$&
$5.6\times 10^{-16}$\\
$0.9$&
$\Omega_\theta$&
$2.7299110395017517\times 10^{-2}$&
$2.7299110395017506\times 10^{-2}$&
$4.1\times 10^{-16}$\\
$0.9$&
$\Omega_\phi$&
$3.0550463796964692\times 10^{-2}$&
$3.0550463796964682\times 10^{-2}$&
$3.6\times 10^{-16}$\\
\mr
$1$&
$\Omega_r$&
$1.9343466898960462\times 10^{-2}$&
$1.9343466898960444\times 10^{-2}$&
$7.8\times 10^{-16}$\\
$1$&
$\Omega_\theta$&
$2.6337035996626332\times 10^{-2}$&
$2.6337035996626321\times 10^{-2}$&
$4.3\times 10^{-16}$\\
$1$&
$\Omega_\phi$&
$2.9662029663040452\times 10^{-2}$&
$2.9662029663040452\times 10^{-2}$&
$4.0\times 10^{-17}$\\
\br
\end{tabular}
}
\end{center}
\end{table}
\section{Analytical solutions of bound geodesics}
\label{sec:ana_orbit}
In this section, we derive the analytical expressions for 
bound geodesic orbits, $r(\lambda)$, $\cos\theta(\lambda)$, $t(\lambda)$ and 
$\phi(\lambda)$, in terms of the elliptic integrals. 
Since we have already derived the orbital frequencies 
in terms of the complete elliptic integrals in \sref{sec:Omega}, 
we can derive the orbits if we replace 
the complete elliptic integrals with the incomplete elliptic integrals. 
However, we have to take account of the initial values of both 
$r$ and $\theta$ and the signs of both $\rmd r/\rmd\lambda$ and 
$\rmd\cos\theta/\rmd\lambda$ at given $\lambda$-time. 
In the following subsections, 
we derive the radial solutions, 
$r(\lambda)$, $t^{(r)}(\lambda)$ and $\phi^{(r)}(\lambda)$,
in \sref{sec:radial_orbits}, 
and the polar solutions, 
$\cos\theta(\lambda)$, $t^{(\theta)}(\lambda)$ and $\phi^{(\theta)}(\lambda)$, 
in \sref{sec:polar_orbits}. 
Finally, we check the consistency of our analytical results 
by comparison with the results of earlier literature in \sref{sec:check}. 
\subsection{Radial solution\,:\,$r(\lambda)$, $t^{(r)}(\lambda)$ and $\phi^{(r)}(\lambda)$}
\label{sec:radial_orbits}
Solving (\ref{eq:ellipt_r_th}) and (\ref{eq:dt_dphi_dlam}), 
we obtain $\lambda(r)$, $t^{(r)}(\lambda)$ and $\phi^{(r)}(\lambda)$ as 
\begin{eqnarray}
\lambda(r) &= \int^r \frac{\rmd r'}{\sqrt{R(r')}},\cr
t^{(r)}(\lambda) &= \int^{r(\lambda)}\frac{T_{r}(r') - \Upsilon_{t^{(r)}}}{\sqrt{R(r')}}\rmd r',\cr
\phi^{(r)}(\lambda) &= \int^{r(\lambda)}\frac{\Phi_{r}(r')-\Upsilon_{\phi^{(r)}}}{\sqrt{R(r')}}\rmd r'.
\label{eq:lam_t_phi_r}
\end{eqnarray}
We derive $r(\lambda)$ inverting $\lambda(r)$. 
Since the period of $r$-motion with respect to $\lambda$ is 
$\Lambda_r=2\pi/\Upsilon_r$, 
we map $\lambda$ to $\lambda^{(r)}$ as 
$\lambda^{(r)}=\lambda-2\pi[\Upsilon_r\lambda/2\pi]/\Upsilon_r$, 
where $[\cdots]$ is the floor function, in the following subsections. 
In order to investigate the integrations in 
(\ref{eq:lam_t_phi_r}) properly, 
we have to take account of $r(\lambda=0)$ 
and the sign of $\rmd r(\lambda)/\rmd\lambda$.
There exist two cases depending on whether the initial value is 
$\rmd r(0)/\rmd\lambda\ge 0$ or $\rmd r(0)/\rmd\lambda\le 0$.
In the following subsections, we consider the two cases separately. 
We note that the expressions of both 
$t^{(r)}(\lambda)$ and $\phi^{(r)}(\lambda)$ in the following subsections 
are valid when $|a|\neq M$. We show 
$t^{(r)}(\lambda)$ and $\phi^{(r)}(\lambda)$ when $|a|=M$ in \ref{sec:a_M}.
\subsubsection{$\rmd r(0)/\rmd\lambda\ge 0$ case}
In this subsection, we consider the case that the initial value of $r(\lambda)$ 
satisfy $\rmd r(0)/\rmd\lambda\ge 0$. We set $r(\lambda=0)=r^{(1)}_0$ in this subsection.
Then $\lambda(r)$ in (\ref{eq:lam_t_phi_r}) can be expressed as
\begin{eqnarray}
\lambda^{(r)}(r) 
&=\int_{r^{(1)}_0}^{r}\frac{\rmd r'}{\sqrt{R(r')}},&\cr
&=\left[\int^r_{r_2}-\int^{r^{(1)}_0}_{r_2}\right]\frac{\rmd r'}{\sqrt{R(r')}}\qquad &r:r^{(1)}_0\rightarrow r_1,\cr
&=\left[-\int^r_{r_2}+2\int^{r_1}_{r_2}-\int^{r^{(1)}_0}_{r_2}\right]\frac{\rmd r'}{\sqrt{R(r')}}\qquad &r:r_1\rightarrow r_2,\cr
&=\left[\int^r_{r_2}+2\int^{r_1}_{r_2}-\int^{r^{(1)}_0}_{r_2}\right]\frac{\rmd r'}{\sqrt{R(r')}}\qquad &r:r_2\rightarrow r^{(1)}_0.
\label{eq:lam_r_r1_decom}
\end{eqnarray}

Thus we find the solution as
\begin{eqnarray}
  \lambda^{(r)}(r) =
  \left\{\begin{array}{ll} 
  \lambda^{(r)}_0(r)-\Lambda_r^{(1)}    \qquad&r:r^{(1)}_0\rightarrow r_1,\cr
  -\lambda^{(r)}_0(r)+\Lambda_r-\Lambda_r^{(1)}    \qquad &r:r_1\rightarrow r_2,\cr
  \lambda^{(r)}_0(r)+\Lambda_r-\Lambda_r^{(1)}    \qquad &r:r_2\rightarrow r^{(1)}_0,
  \end{array}\right.
\label{eq:lam_r_r1}
\end{eqnarray}
where
\begin{eqnarray}
 \lambda^{(r)}_0(r) = \frac{1}{\sqrt{1-\C{E}^2}}\frac{2}{\sqrt{(r_1-r_3)(r_2-r_4)}}
F\left(\arcsin y_r,k_r\right),
\end{eqnarray}
and $\Lambda_r^{(1)}=\lambda^{(r)}_0(r^{(1)}_0)$.

Inverting \eref{eq:lam_r_r1}, we derive $r(\lambda)$ as 
\begin{eqnarray}
r(\lambda)=\frac{r_3(r_1-r_2) {\rm sn}^2(u_r(\lambda),k_r)-r_2(r_1-r_3)}
{(r_1-r_2) {\rm sn}^2(u_r(\lambda),k_r)-(r_1-r_3)},
\label{eq:ana_r}
\end{eqnarray}
where 
${\rm sn}(u,k)$ is Jacobi's elliptic function which is defined as 
the inverse function of the incomplete elliptic integrals, 
$u=F(\varphi,k)$, and 
\begin{eqnarray}
\fl
u_r(\lambda)=
  \left\{\begin{array}{ll} 
  2K(k_r)(\lambda^{(r)}+\Lambda_r^{(1)})/\Lambda_r \qquad &(0\le \lambda^{(r)} \le \Lambda_r/2-\Lambda_r^{(1)}),\\
  2K(k_r)(-\lambda^{(r)}+\Lambda_r-\Lambda_r^{(1)})/\Lambda_r \qquad &(\Lambda_r/2-\Lambda_r^{(1)}\le \lambda^{(r)} \le \Lambda_r-\Lambda_r^{(1)}),\\
  2K(k_r)(\lambda^{(r)}-\Lambda_r+\Lambda_r^{(1)})/\Lambda_r \qquad &(\Lambda_r-\Lambda_r^{(1)}\le \lambda^{(r)} \le \Lambda_r).
  \end{array}\right.
\end{eqnarray}

Combining the results of (\ref{eq:lam_r_r1_decom}) with  
the results of \ref{sec:integral_formula}, 
we can derive $t^{(r)}$ and $\phi^{(r)}$ in (\ref{eq:lam_t_phi_r}) as 
\begin{eqnarray}
\fl
t^{(r)} =
\frac{2}{\sqrt{(1-\C{E}^2)(r_1-r_3)(r_2-r_4)}}\cr
\times\left\{\frac{\C{E}}{2}\left[
(r_2-r_3)(r_1+r_2+r_3+r_4)\tilde{\Pi}_r(\psi_r,-h_r,k_r)\right.\right.\cr
+\left.(r_1-r_3)(r_2-r_4)\tilde{E}_r(\psi_r,h_r,k_r)\right]\cr
+2M\C{E}(r_2-r_3)\tilde{\Pi}_r(\psi_r,-h_r,k_r)\cr
-\frac{2M}{r_{+}-r_{-}}\left[\frac{(4M^2\C{E}-a\C{L}_{z})r_{+}-2Ma^2\C{E}}{r_3-r_{+}}
\frac{r_2-r_3}{r_2-r_{+}}\tilde{\Pi}_r(\psi_r,-h_{+},k_r)\right.\cr
\left.\left.-(+\leftrightarrow -)\right]\right\},\cr
\fl
\phi^{(r)} = 
-\frac{2a}{(r_{+}-r_{-})\sqrt{(1-\C{E}^2)(r_1-r_3)(r_2-r_4)}}
\left[\frac{(2M\C{E}r_{+}-a\C{L}_{z})(r_2-r_3)}{(r_3-r_{+})(r_2-r_{+})}
\tilde{\Pi}_r(\psi_r,-h_{+},k_r)\right.\cr
-\left.(+\leftrightarrow -)\right],
\label{eq:ana_tr_phir}
\end{eqnarray}
where 
$\psi_r=\arcsin[{\rm sn}(u_r,k_r)]$, 
$\tilde{E}_r(\psi_r,c,k_r)=E_r(\psi_r,c,k_r)-\frac{\Upsilon_r\lambda^{(r)}}{\pi}E(k_r)$, 
$\tilde{\Pi}_r(\psi_r,c,k_r)=\Pi_r(\psi_r,c,k_r)-\frac{\Upsilon_r\lambda^{(r)}}{\pi}\Pi(c,k_r)$ and 

\begin{eqnarray}
E_r^{(0)}(\psi_r,c,k_r)&=&E(\psi_r,k_r)+\frac{\sin\psi_r\sqrt{(1-\sin^2\psi_r)(1-k_{r}^{2}\sin^2\psi_r)}}{\sin^2\psi_r-c^{-1}},\cr
E_r(\psi_r,c,k_r)&=&E_r^{(0)}(\psi_r,c,k_r)-E_r^{(0)}(\psi_r(0),c,k_r)\cr
&&\qquad\qquad\qquad\qquad {\rm for}\qquad 0\le \lambda^{(r)} \le \Lambda_r/2-\Lambda_r^{(1)},\cr
&=&-E_r^{(0)}(\psi_r,c,k_r)+2E(k_r)-E_r^{(0)}(\psi_r(0),c,k_r)\cr
&&\qquad\qquad\qquad\qquad {\rm for}\qquad \Lambda_r/2-\Lambda_r^{(1)}\le \lambda^{(r)} \le \Lambda_r-\Lambda_r^{(1)},\cr
&=&E_r^{(0)}(\psi_r,c,k_r)+2E(k_r)-E_r^{(0)}(\psi_r(0),c,k_r)\cr
&&\qquad\qquad\qquad\qquad {\rm for}\qquad  \Lambda_r-\Lambda_r^{(1)}\le \lambda^{(r)} \le \Lambda_r,\cr
\Pi_r(\psi_r,c,k_r)&=&\Pi(\psi_r,c,k_r)-\Pi(\psi_r(0),c,k_r)\cr
&&\qquad\qquad\qquad\qquad {\rm for}\qquad 0\le \lambda^{(r)} \le \Lambda_r/2-\Lambda_r^{(1)},\cr
&=&-\Pi(\psi_r,c,k_r)+2\Pi(c,k_r)-\Pi(\psi_r(0),c,k_r)\cr
&&\qquad\qquad\qquad\qquad {\rm for}\qquad \Lambda_r/2-\Lambda_r^{(1)}\le \lambda^{(r)} \le \Lambda_r-\Lambda_r^{(1)},\cr
&=&\Pi(\psi_r,c,k_r)+2\Pi(c,k_r)-\Pi(\psi_r(0),c,k_r)\cr
&&\qquad\qquad\qquad\qquad {\rm for}\qquad \Lambda_r-\Lambda_r^{(1)}\le \lambda^{(r)} \le \Lambda_r.
\end{eqnarray}
\subsubsection{$\rmd r(0)/\rmd\lambda\le 0$ case}
In this subsection, we consider the case that the initial value of $r(\lambda)$ 
satisfy $\rmd r(0)/\rmd\lambda\le 0$. We set $r(\lambda=0)=r^{(2)}_0$ in this subsection.
Then $\lambda(r)$ in (\ref{eq:lam_t_phi_r}) can be expressed as
\begin{eqnarray}
\lambda^{(r)}(r) 
&=\int_{r^{(2)}_0}^{r}\frac{\rmd r'}{\sqrt{R(r')}},&\cr
&=\left[-\int^r_{r_2}+\int^{r^{(2)}_0}_{r_2}\right]\frac{\rmd r'}{\sqrt{R(r')}} &\qquad r:r^{(2)}_0\rightarrow r_2,\cr
&=\left[\int^r_{r_2}+\int^{r^{(2)}_0}_{r_2}\right]\frac{\rmd r'}{\sqrt{R(r')}}&\qquad r:r_2\rightarrow r_1,\cr
&=\left[-\int^r_{r_2}+2\int^{r_1}_{r_2}+\int^{r^{(2)}_0}_{r_2}\right]\frac{\rmd r'}{\sqrt{R(r')}}&\qquad r:r_1\rightarrow r^{(2)}_0.
\label{eq:lam_r_r2_decom}
\end{eqnarray}

Thus we find the solution as
\begin{eqnarray}
  \lambda^{(r)}(r) =
  \left\{\begin{array}{ll} 
  -\lambda^{(r)}_0(r)+\Lambda_r^{(2)}    \qquad &r:r^{(2)}_0\rightarrow r_2,\\
  \lambda^{(r)}_0(r)+\Lambda_r^{(2)}    \qquad &r:r_2\rightarrow r_1,\\
  -\lambda^{(r)}_0(r)+\Lambda_r+\Lambda_r^{(2)}    \qquad &r:r_1\rightarrow r^{(2)}_0,
  \end{array}\right.
\label{eq:lam_r_r2}
\end{eqnarray}
where $\Lambda_r^{(2)}=\lambda^{(r)}_0(r^{(2)}_0)$.

Then we obtain $r(\lambda)$ in the same form in (\ref{eq:ana_r})
inverting (\ref{eq:lam_r_r2}), 
but $u_r(\lambda)$ in (\ref{eq:ana_r}) is modified as
\begin{eqnarray}
\fl
  u_r(\lambda)=
  \left\{\begin{array}{ll} 
  2K(k_r)(-\lambda^{(r)}+\Lambda_r^{(2)})/\Lambda_r \qquad &(0\le \lambda^{(r)} \le \Lambda_r^{(2)}),\\
  2K(k_r)(\lambda^{(r)}-\Lambda_r^{(2)})/\Lambda_r \qquad &(\Lambda_r^{(2)}\le \lambda^{(r)} \le \Lambda_r/2+\Lambda_r^{(2)}),\\
  2K(k_r)(-\lambda^{(r)}+\Lambda_r+\Lambda_r^{(2)})/\Lambda_r \qquad &(\Lambda_r/2+\Lambda_r^{(2)}\le \lambda^{(r)} \le \Lambda_r).
  \end{array}\right.
\end{eqnarray}

Combining the results of (\ref{eq:lam_r_r2_decom}) with  
the results of \ref{sec:integral_formula}, 
we can derive $t^{(r)}$ and $\phi^{(r)}$ 
in the same form in (\ref{eq:ana_tr_phir}), 
but $E_r(\psi_r,c,k_r)$ and $\Pi_r(\psi_r,c,k_r)$ are modified as 
\begin{eqnarray}
  E_r(\psi_r,c,k_r)&=&
  -E_r^{(0)}(\psi_r,c,k_r)+E_r^{(0)}(\psi_r(0),c,k_r)\cr
&&\qquad\qquad\qquad\qquad {\rm for}\qquad 0\le \lambda^{(r)} \le \Lambda_r^{(2)},\cr
&=&E_r^{(0)}(\psi_r,c,k_r)+E_r^{(0)}(\psi_r(0),c,k_r)\cr
&&\qquad\qquad\qquad\qquad {\rm for}\qquad \Lambda_r^{(2)}\le \lambda^{(r)} \le \Lambda_r/2+\Lambda_r^{(2)},\cr
&=&-E_r^{(0)}(\psi_r,c,k_r)+2E(k_r)+E_r^{(0)}(\psi_r(0),c,k_r)\cr
&&\qquad\qquad\qquad\qquad {\rm for}\qquad  \Lambda_r/2+\Lambda_r^{(2)}\le \lambda^{(r)} \le \Lambda_r,\cr
  \Pi_r(\psi_r,c,k_r)&=&
  -\Pi(\psi_r,c,k_r)+\Pi(\psi_r(0),c,k_r)\cr
&&\qquad\qquad\qquad\qquad {\rm for}\qquad  0\le \lambda^{(r)} \le \Lambda_r^{(2)},\cr
&=&\Pi(\psi_r,c,k_r)+\Pi(\psi_r(0),c,k_r)\cr
&&\qquad\qquad\qquad\qquad {\rm for}\qquad  \Lambda_r^{(2)}\le \lambda^{(r)} \le \Lambda_r/2+\Lambda_r^{(2)},\cr
&=&-\Pi(\psi_r,c,k_r)+2\Pi(c,k_r)+\Pi(\psi_r(0),c,k_r)\cr
&&\qquad\qquad\qquad\qquad {\rm for}\qquad  \Lambda_r/2+\Lambda_r^{(2)}\le \lambda^{(r)} \le \Lambda_r.
\end{eqnarray}
\subsection{Polar Solution\,:\,$\cos\theta(\lambda)$, $t^{(\theta)}(\lambda)$ and $\phi^{(\theta)}(\lambda)$}
\label{sec:polar_orbits}
Solving (\ref{eq:ellipt_r_th}) and (\ref{eq:dt_dphi_dlam}), 
we obtain $\lambda(\cos\theta)$, $t^{(\theta)}(\lambda)$ and 
$\phi^{(\theta)}(\lambda)$ as 
\begin{eqnarray}
\lambda(\theta)&=
\int^{\cos\theta}\frac{\rmd\cos\theta '}{\sqrt{\Theta(\cos\theta ')}},\cr
t^{(\theta)}(\lambda) &= \int^{\cos\theta}\frac{T_{\theta}(\cos\theta ') -\Upsilon_{t^{(\theta)}}}{\sqrt{\Theta(\cos\theta ')}}\rmd\cos\theta ',\cr
\phi^{(\theta)}(\lambda) &= \int^{\cos\theta}\frac{\Phi_{\theta}(\cos\theta ')-\Upsilon_{\phi^{(\theta)}}}{\sqrt{\Theta(\cos\theta ')}}\rmd\cos\theta '.
\label{eq:lam_t_phi_th}
\end{eqnarray}
We derive $\cos\theta(\lambda)$ inverting $\lambda(\theta)$. 
Since the period of $\theta$-motion with respect to $\lambda$ is 
$\Lambda_\theta=2\pi/\Upsilon_\theta$, 
we map $\lambda$ to $\lambda^{(\theta)}$ as 
$\lambda^{(\theta)}=\lambda-2\pi[\Upsilon_\theta\lambda/2\pi]/\Upsilon_\theta$ 
in the following subsections. 
In order to investigate the integrations in 
(\ref{eq:lam_t_phi_th}) properly, 
we have to take account of $\cos\theta(\lambda=0)$ 
and the sign of $\rmd\cos\theta(\lambda)/\rmd\lambda$.
There exist two cases depending on whether  the initial value is 
$\rmd\cos\theta(0)/\rmd\lambda\ge 0$ or $\rmd\cos\theta(0)/\rmd\lambda\le 0$.
In the following subsections, we consider the two cases separately. 
\subsubsection{$d\cos\theta(0)/d\lambda\ge 0$ case}
In this subsection, we consider the case that the initial value of $\cos\theta(\lambda)$ 
satisfies $\rmd\cos\theta(0)/\rmd\lambda\ge 0$. 
We set $\theta(\lambda=0)=\theta^{(1)}_0$ in this subsection.
Then $\lambda(\theta)$ in (\ref{eq:lam_t_phi_th}) can be expressed as
\begin{eqnarray}
\fl
\lambda^{(\theta)}(\theta) 
&=\int_{\cos\theta^{(1)}_0}^{\cos\theta}\frac{\rmd\cos\theta '}{\sqrt{\Theta(\cos\theta ')}},&\cr
&=\left[\int^{\cos\theta}_{0}-\int^{\cos\theta^{(1)}_0}_{0}\right]\frac{\rmd\cos\theta '}{\sqrt{\Theta(\cos\theta ')}}\qquad &\theta:\theta^{(1)}_0\rightarrow \theta_{\rm min},\cr
&=\left[-\int^{\cos\theta}_{0}+2\int^{\cos\theta_{\rm min}}_{0}-\int^{\cos\theta^{(1)}_0}_{0}\right]\frac{\rmd\cos\theta '}{\sqrt{\Theta(\cos\theta ')}}\qquad &\theta:\theta_{\rm min}\rightarrow \pi-\theta_{\rm min},\cr
&=\left[\int^{\cos\theta}_{0}+4\int^{\cos\theta_{\rm min}}_{0}-\int^{\cos\theta^{(1)}_0}_{0}\right]\frac{\rmd\cos\theta '}{\sqrt{\Theta(\cos\theta ')}}\qquad &\theta:\pi-\theta_{\rm min}\rightarrow \theta^{(1)}_0.
\label{eq:lam_th_th1_decom}
\end{eqnarray}

Thus we find the solution as
\begin{eqnarray}
  \lambda^{(\theta)}(\theta) =
  \left\{\begin{array}{ll} 
  \lambda^{(\theta)}_0(\theta)-\Lambda_\theta^{(1)}\qquad &\theta:\theta^{(1)}_0\rightarrow \theta_{\rm min},\\
  -\lambda^{(\theta)}_0(\theta)+\Lambda_\theta/2-\Lambda_\theta^{(1)} \qquad &\theta:\theta_{\rm min}\rightarrow \pi-\theta_{\rm min},\\
  \lambda^{(\theta)}_0(\theta)+\Lambda_{\theta}-\Lambda_\theta^{(1)} \qquad &\theta:\pi-\theta_{\rm min}\rightarrow \theta^{(1)}_0,
  \end{array}\right.
\label{eq:lam_th_th1}
\end{eqnarray}
where 
\begin{eqnarray}
 \lambda^{(\theta)}_0(\theta) = \frac{1}{\C{L}_{z}\sqrt{\epsilon_0 z_{+}}}
 F\left(\arcsin y_\theta,k_\theta\right),
\end{eqnarray}
and $\Lambda_\theta^{(1)}=\lambda^{(\theta)}_0(\theta^{(1)}_0)$.

Inverting (\ref{eq:lam_th_th1}), we derive $\cos\theta(\lambda)$ as
\begin{eqnarray}
\cos\theta(\lambda) = \sqrt{z_{-}}{\rm sn}(u_\theta(\lambda),k_\theta),
\label{eq:ana_th}
\end{eqnarray}
where
\begin{eqnarray*}
\fl
  u_\theta(\lambda)=
  \left\{\begin{array}{ll} 
  4K(k_\theta)(\lambda^{(\theta)}+\Lambda_\theta^{(1)})/\Lambda_\theta \qquad &(0\le \lambda^{(\theta)} \le \Lambda_\theta/4-\Lambda_\theta^{(1)}),\\
  4K(k_\theta)(-\lambda^{(\theta)}+\Lambda_\theta/2-\Lambda_\theta^{(1)})/\Lambda_\theta \qquad &(\Lambda_\theta/4-\Lambda_\theta^{(1)}\le \lambda^{(\theta)} \le 3\Lambda_\theta/4-\Lambda_\theta^{(1)}),\\
  4K(k_\theta)(\lambda^{(\theta)}-\Lambda_\theta+\Lambda_\theta^{(1)})/\Lambda_\theta \qquad &(3\Lambda_\theta/4-\Lambda_\theta^{(1)}\le \lambda^{(\theta)} \le \Lambda_\theta).
  \end{array}\right.
\end{eqnarray*}

Using the results of (\ref{eq:lam_th_th1_decom}), 
we derive $t^{(\theta)}$ and $\phi^{(\theta)}$ as
\begin{eqnarray}
t^{(\theta)} =& 
\frac{a^2\C{E}z_{+}}{\C{L}_{z}\sqrt{\epsilon_0 z_{+}}}
\left[\frac{2\Upsilon_\theta\lambda^{(\theta)}}{\pi}
E(k_\theta)-E_\theta(\psi_\theta,k_\theta)\right],\cr
\phi^{(\theta)} =& \frac{1}{\sqrt{\epsilon_0 z_{+}}}
\left[\Pi_\theta(\psi_\theta,-z_{-},k_\theta)
-\frac{2\Upsilon_\theta\lambda^{(\theta)}}{\pi}\Pi(-z_{-},k_\theta)\right],
\label{eq:ana_tth_phith}
\end{eqnarray}
where $\psi_\theta=\arcsin[{\rm sn}(u_\theta,k_\theta)]$ and 
\begin{eqnarray}
\fl
  E_\theta(\psi_\theta,k_\theta)&=
  E(\psi_\theta,k_\theta)-E(\psi_\theta(0),k_\theta)\cr
&\qquad\qquad\qquad\qquad {\rm for}\qquad 0\le \lambda^{(\theta)} \le \Lambda_\theta/4-\Lambda_\theta^{(1)},\cr
&=-E(\psi_\theta,k_\theta)+2E(k_\theta)-E(\psi_\theta(0),k_\theta)\cr
&\qquad\qquad\qquad\qquad {\rm for}\qquad \Lambda_\theta/4-\Lambda_\theta^{(1)}\le \lambda^{(\theta)} \le 3\Lambda_\theta/4-\Lambda_\theta^{(1)},\cr
&=E(\psi_\theta,k_\theta)+4E(k_\theta)-E(\psi_\theta(0),k_\theta)\cr
&\qquad\qquad\qquad\qquad {\rm for}\qquad 3\Lambda_\theta/4-\Lambda_\theta^{(1)}\le \lambda^{(\theta)} \le \Lambda_\theta,
\end{eqnarray}
\begin{eqnarray}
\fl
  \Pi_\theta(\psi_\theta,c,k_\theta)&=
  \Pi(\psi_\theta,c,k_\theta)-\Pi(\psi_\theta(0),c,k_\theta)\cr
&\qquad\qquad\qquad\qquad {\rm for}\qquad 0\le \lambda^{(\theta)} \le \Lambda_\theta/4-\Lambda_\theta^{(1)},\cr
&=-\Pi(\psi_\theta,c,k_\theta)+2\Pi(c,k_\theta)-\Pi(\psi_\theta(0),c,k_\theta)\cr
&\qquad\qquad\qquad\qquad {\rm for}\qquad \Lambda_\theta/4-\Lambda_\theta^{(1)}\le \lambda^{(\theta)} \le 3\Lambda_\theta/4-\Lambda_\theta^{(1)},\cr
&=\Pi(\psi_\theta,c,k_\theta)+4\Pi(c,k_\theta)-\Pi(\psi_\theta(0),c,k_\theta)\cr
&\qquad\qquad\qquad\qquad {\rm for}\qquad 3\Lambda_\theta/4-\Lambda_\theta^{(1)}\le \lambda^{(\theta)} \le \Lambda_\theta.
\end{eqnarray}
\subsubsection{$\rmd\cos\theta(0)/\rmd\lambda\le 0$ case}
In this subsection, we consider the case that the initial value of $\cos\theta(\lambda)$ 
satisfies $\rmd\cos\theta(0)/\rmd\lambda\le 0$. 
We set $\theta(\lambda=0)=\theta^{(2)}_0$ in this subsection.
Then $\lambda(\theta)$ in (\ref{eq:lam_t_phi_th}) can be expressed as
\begin{eqnarray}
\fl
\lambda^{(\theta)}(\theta) 
&=\int_{\cos\theta^{(2)}_0}^{\cos\theta}\frac{\rmd\cos\theta '}{\sqrt{\Theta(\cos\theta ')}},&\cr
&=\left[-\int^{\cos\theta}_{0}+\int^{\cos\theta^{(2)}_0}_{0}\right]\frac{\rmd\cos\theta '}{\sqrt{\Theta(\cos\theta ')}}\qquad &\theta:\theta^{(2)}_0\rightarrow \pi-\theta_{\rm min},\cr
&=\left[\int^{\cos\theta}_{0}+2\int^{\cos\theta_{\rm min}}_{0}+\int^{\cos\theta^{(2)}_0}_{0}\right]\frac{\rmd\cos\theta '}{\sqrt{\Theta(\cos\theta ')}}\qquad &\theta:\pi-\theta_{\rm min}\rightarrow \theta_{\rm min},\cr
&=\left[-\int^{\cos\theta}_{0}+4\int^{\cos\theta_{\rm min}}_{0}+\int^{\cos\theta^{(2)}_0}_{0}\right]\frac{\rmd\cos\theta '}{\sqrt{\Theta(\cos\theta ')}}\qquad &\theta:\theta_{\rm min}\rightarrow \theta^{(2)}_0.
\label{eq:lam_th_th2_decom}
\end{eqnarray}

Thus we find the solution as
\begin{eqnarray}
  \lambda^{(\theta)}(\theta) =
  \left\{\begin{array}{ll} 
  -\lambda^{(\theta)}_0(\theta)+\Lambda_\theta^{(2)}\qquad &\theta:\theta^{(2)}_0\rightarrow \pi-\theta_{\rm min},\\
  \lambda^{(\theta)}_0(\theta)+\Lambda_\theta/2+\Lambda_\theta^{(2)} \qquad &\theta:\pi-\theta_{\rm min}\rightarrow \theta_{\rm min},\\
  -\lambda^{(\theta)}_0(\theta)+\Lambda_{\theta}+\Lambda_\theta^{(2)}\qquad &\theta:\theta_{\rm min}\rightarrow \theta^{(2)}_0,
  \end{array}\right.
\label{eq:lam_th_th2}
\end{eqnarray}
where $\Lambda_\theta^{(2)}=\lambda^{(\theta)}_0(\theta^{(2)}_0)$.

Then we obtain $\cos\theta(\lambda)$ in the same form in (\ref{eq:ana_th})
inverting (\ref{eq:lam_th_th2}), 
but $u_\theta(\lambda)$ in (\ref{eq:ana_th}) is modified as
\begin{eqnarray*}
\fl
  u_\theta(\lambda)=
  \left\{\begin{array}{ll} 
  4K(k_\theta)(-\lambda^{(\theta)}+\Lambda_\theta^{(2)})/\Lambda_\theta \qquad &(0\le \lambda^{(\theta)} \le \Lambda_\theta/4+\Lambda_\theta^{(2)}),\\
  4K(k_\theta)(\lambda^{(\theta)}-\Lambda_\theta/2-\Lambda_\theta^{(2)})/\Lambda_\theta \qquad &(\Lambda_\theta/4+\Lambda_\theta^{(2)}\le \lambda^{(\theta)} \le 3\Lambda_\theta/4+\Lambda_\theta^{(2)}),\\
  4K(k_\theta)(-\lambda^{(\theta)}+\Lambda_\theta+\Lambda_\theta^{(2)})/\Lambda_\theta \qquad &(3\Lambda_\theta/4+\Lambda_\theta^{(2)}\le \lambda^{(\theta)} \le \Lambda_\theta).
  \end{array}\right.
\end{eqnarray*}

Using the results of (\ref{eq:lam_th_th2_decom}), 
we derive $t^{(\theta)}$ and $\phi^{(\theta)}$ as
in the same form in (\ref{eq:ana_tth_phith}), 
but $E_\theta(\psi_\theta,k_\theta)$ and 
$\Pi_\theta(\psi_\theta,c,k_\theta)$ are modified as 
\begin{eqnarray}
\fl
  E_\theta(\psi_\theta,k_\theta)&=
  -E(\psi_\theta,k_\theta)+E(\psi_\theta(0),k_\theta)\cr
&\qquad\qquad\qquad\qquad {\rm for}\qquad 0\le \lambda^{(\theta)} \le \Lambda_\theta/4+\Lambda_\theta^{(2)},\cr
&=  E(\psi_\theta,k_\theta)+2E(k_\theta)+E(\psi_\theta(0),k_\theta)\cr
&\qquad\qquad\qquad\qquad {\rm for}\qquad \Lambda_\theta/4+\Lambda_\theta^{(2)}\le \lambda^{(\theta)} \le 3\Lambda_\theta/4+\Lambda_\theta^{(2)},\cr
&=  -E(\psi_\theta,k_\theta)+4E(k_\theta)+E(\psi_\theta(0),k_\theta)\cr
&\qquad\qquad\qquad\qquad {\rm for}\qquad 3\Lambda_\theta/4+\Lambda_\theta^{(2)}\le \lambda^{(\theta)} \le \Lambda_\theta,
\end{eqnarray}
\begin{eqnarray}
\fl
  \Pi_\theta(\psi_\theta,c,k_\theta)&=
  -\Pi(\psi_\theta,c,k_\theta)+\Pi(\psi_\theta(0),c,k_\theta)\cr
&\qquad\qquad\qquad\qquad {\rm for}\qquad 0\le \lambda^{(\theta)} \le \Lambda_\theta/4+\Lambda_\theta^{(2)},\cr
&=  \Pi(\psi_\theta,c,k_\theta)+2\Pi(c,k_\theta)+\Pi(\psi_\theta(0),c,k_\theta)\cr
&\qquad\qquad\qquad\qquad {\rm for}\qquad \Lambda_\theta/4+\Lambda_\theta^{(2)}\le \lambda^{(\theta)} \le 3\Lambda_\theta/4+\Lambda_\theta^{(2)},\cr
&= -\Pi(\psi_\theta,c,k_\theta)+4\Pi(c,k_\theta)+\Pi(\psi_\theta(0),c,k_\theta)\cr
&\qquad\qquad\qquad\qquad {\rm for}\qquad 3\Lambda_\theta/4+\Lambda_\theta^{(2)}\le \lambda^{(\theta)} \le \Lambda_\theta.
\end{eqnarray}
\subsection{Consistency check of the analytical solution}
\label{sec:check}
In this subsection, we compare the analytical results of bound geodesics, 
$t(\lambda)$, $r(\lambda)$, $\cos\theta(\lambda)$ and $\phi(\lambda)$, 
with earlier literature as a consistency check. 
We can compare $r(\lambda)$ in \eref{eq:ana_r} with that in 
\cite{Dexter,Kraniotis:2007}, 
in which the integral of motion, 
$\int\rmd r/\sqrt{R(r)}=\int\rmd\cos\theta/\sqrt{\Theta(\cos\theta)}$, 
is solved in terms of Jacobi's elliptic function using $\theta$ as 
the independent variable. 
Since \cite{Dexter} deals with 
null geodesics and \cite{Kraniotis:2007} does not give explicit expressions 
including the turning points, we can not compare them exactly. 
However, we find that the formal expressions of $r(\lambda)$ 
in this paper and \cite{Dexter,Kraniotis:2007} are consistent. 
We can also compare the formal expression of $\cos\theta(\lambda)$ 
in \eref{eq:ana_th} with that in 
\cite{Dexter}, in which the integral of motion 
is solved in terms of Jacobi's elliptic function using $r$ as 
the independent variable. Then we find that the formal expressions of 
$\cos\theta(\lambda)$ in this paper and \cite{Dexter} are consistent. 
Although \cite{Dexter} derived $t$ and $\phi$ in terms of 
Carlson elliptic integrals \cite{carlson1988}, 
it seems difficult to compare the expressions of 
both $t$ and $\phi$ in this paper with that of null geodesics 
in \cite{Dexter}. 
Thus we compare the expressions of each integrals such as 
$\int^r_{r_2} (r')^2\rmd r'/\sqrt{R(r')}$ in \ref{sec:integral_formula} 
with that in \cite{carlson1988}, in which formulas for 
$\int_y^x \prod_{i=1}^{5} (a_i+b_i t)^{p_i/2} \rmd t$ 
are derived in terms of Carlson elliptic integrals, 
where all quantities are real, $x>y$ and $a_i+b_i t > 0$ for $y < t < x$. 
Using transformation of Carlson elliptic integrals 
of the third kind \cite{carlson1970}, it is straightforward to check 
$\int^r_{r_2} (r')^2\rmd r'/\sqrt{R(r')}$ in \ref{sec:integral_formula} 
agrees with the corresponding formula in \cite{carlson1988}. We can also 
check the other integrals in this paper agree with that in \cite{carlson1988}. 
These analytical checks show that the expressions in this paper 
are consistent. 

Moreover, we compare the analytical results in this paper with 
that of numerical integration method. It is a good check on the results 
that both the coefficients and turning points in each integrals 
are consistent. 
In numerical integration of geodesic equation, 
as explained in \sref{sec:geodesics}, 
we transform $r$ as $r=pM/(1+e\cos\psi)$ and 
$\cos\theta$ as $\cos\theta=\cos\theta_{\rm min}\cos\chi$ respectively. 
Using (\ref{eq:geodesic_lam}), 
we obtain the following set of differential equations~\cite{Drasco:2004}
\begin{eqnarray}
\fl
\frac{\rmd\psi}{\rmd\lambda} = \frac{M\sqrt{(1-\C{E}^2)\{(p-p_3)-e(p+p_3\cos\psi)\}\{(p-p_4)+e(p-p_4\cos\psi)\}}}{1-e^2},\cr
\fl
\frac{\rmd\chi}{\rmd\lambda} = \sqrt{a^2(1-\C{E}^2)(z_{+}-z_{-}\cos^2\chi)},\cr
\fl
\frac{\rmd t}{\rmd\lambda} = T_{\rm r}(r)+T_\theta(\cos\theta) + a\C{L}_{z},\cr
\fl
\frac{\rmd\phi}{\rmd\lambda} = \Phi_{\rm r}(r)+\Phi_\theta(\cos\theta) - a\C{E},
\label{eq:numerical_geodesic_lam}
\end{eqnarray}
where $p_3=r_3(1-e)/M$ and $p_4=r_4(1+e)/M$. 
We can numerically solve (\ref{eq:numerical_geodesic_lam}) accurately 
without taking account of the turning points of both $r$ and
 $\cos\theta$ because 
both $\psi$ and $\chi$ are monotonic increasing functions of time.

\begin{figure}[htbp]
\begin{center}
\includegraphics[scale=.8]{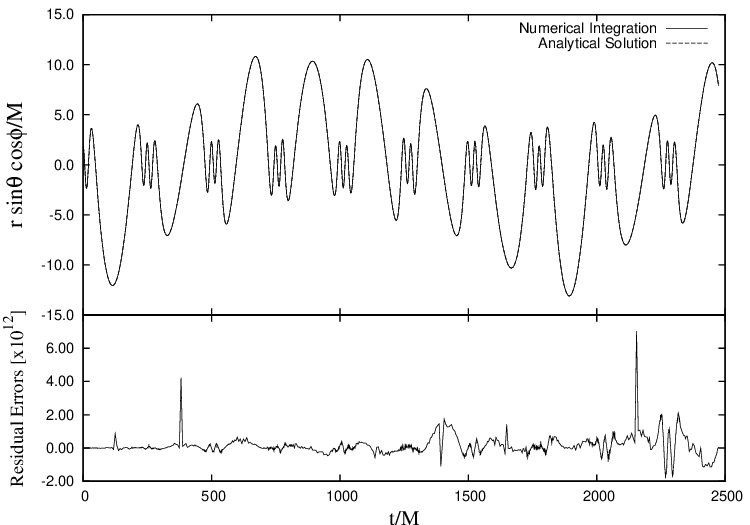}
\end{center}
\caption{
Comparison of the function $x(t)=r(t)\sin\theta(t)\cos\phi(t)$ computed
using  the analytical expressions of this paper with
the result of a numerical integration. 
In this figure, we set orbital elements as 
$a=0.9M$, $p=4M$, $e=0.7$ and $\theta_{\rm inc}=40^{\circ}$. 
And we set initial value of $\psi$ and $\chi$ as $\psi(0)=0$ and $\chi(0)=0$ 
respectively, which correspond to $r(0)=r_0^{(1)}=pM/(1+e)$ and 
$\theta(0)=\theta_0^{(1)}=\pi/2$. 
Upper figure shows plots of both analytical solution, $x_{\rm A}(t)$, 
and numerical integration method, $x_{\rm N}(t)$. 
Lower figure shows the residual errors between 
the results of both analytical solution and numerical integration method, 
$x_{\rm A}(t)-x_{\rm N}(t)$. 
}
\label{fig:orbit_check}
\end{figure}

In \fref{fig:orbit_check}, we compare the results of 
our analytical expressions with the results from the numerical integration
 method. 
We choose orbital parameters as $a=0.9M$, $p=4M$, $e=0.7$ and
 $\theta_{\rm inc}=40^{\circ}$. 
And  set the initial values of $\psi$ and $\chi$ as $\psi(0)=0$ and
 $\chi(0)=0$ 
respectively, which correspond to $r(0)=r_0^{(1)}=pM/(1+e)$ and 
$\theta(0)=\theta_0^{(1)}=\pi/2$.
For numerical integration of (\ref{eq:numerical_geodesic_lam}), 
we use the 4th order Runge-Kutta method with 
non-adaptive step-size control~\cite{Recipes}. 
This figure shows that the analytical solutions of geodesic equation in this paper 
exactly represent the solutions of bound geodesic orbits around a
 Kerr black hole.
\section{Summary}
\label{sec:summary}
We have derived  analytical solutions for bound timelike 
geodesics in Kerr spacetime. 
This is the first time that 
analytical expressions of the fundamental frequencies 
are derived in terms of the elliptic integrals. 
The analytical expressions of the orbits, $(t, r, \cos\theta, \phi)$,  
have been also derived in terms of the elliptic integrals 
using Mino time as the independent variable for the first time. 
Since Mino time decouples the 
$r$ and $\theta$-motion, 
it leads to forms simpler than that in \cite{Dexter} for null case 
if we suitably transform variables, $r$ and $\theta$. 
We checked the consistency of the analytical expressions 
comparing them with the analytical expressions for the other cases, 
post-Newtonian approximation and numerical integration method.

We can apply these solutions to the computation of gravitational waves from 
EMRIs. Gravitational waves from EMRIs are described 
by the Teukolsky formalism~\cite{Teukolsky:1973ha}. 
In the frequency domain calculation of the 
Teukolsky formalism~\cite{chapter,ST}, 
we can use the analytical solutions directory~\cite{Drasco:2006,FHT} 
and compute the orbits more accurately than numerical integration of 
geodesic equation. In principle, we can compute the orbits 
with machine accuracy. 
Using the analytical expressions of radial and polar motion in this paper, 
we showed that the analytical expressions enable one to 
compute gravitational waves from EMRIs very accurately~\cite{FHT}. 
Although it may takes longer time to compute the orbits 
using the analytical solutions than using 
numerical integration method~\cite{Rauch}, but see \cite{Dexter}, 
it is not serious in computing gravitational waves. 
This is because we compute orbits only for one orbital period of 
radial and polar motion, $\Lambda_r$ and $\Lambda_\theta$, and 
computation time of the orbits, $\sim$seconds, 
is much smaller than that of gravitational waves from EMRIs, 
$\sim$hours to days~\cite{Drasco:2006,FHT,Hughes:2005qb}. 
Thus we believe that the analytical solutions are very useful 
for the computation of gravitational waves from EMRIs. 
In the time domain calculation of the Teukolsky formalism
(see brief review in section 3.8 in \cite{Kostas_Review}), 
we may need the inversion of $t(\lambda)$ 
in order to compute the orbits, $r(\lambda)$, $\cos\theta(\lambda)$, 
$t(\lambda)$ and $\phi(\lambda)$, in the coordinate time. 
Although we do not know the analytical expression of $\lambda(t)$, 
we may easily obtain $\lambda(t)$ by numerical iteration 
if we set the initial solution as $\lambda=t/\Gamma$. 
Thus it may  also be useful in the time domain calculation 
if the numerical iteration converges faster than 
the numerical integration of the geodesic equation. 

We may also apply these solutions to investigate the properties of geodesics
of Kerr black holes.
 Although it seems difficult to classify orbits in the strong field 
because of its complexities, Levin \etal 
recently suggested a taxonomy of orbits 
introducing a rational number which is constructed 
from orbital frequencies~\cite{Levin1,Levin2}. 
Both the analytical expressions of the fundamental frequencies in this paper 
and the taxonomy of orbits may help us to discuss 
the conditions characterizing zoom-whirl orbits~\cite{GK} and 
 other extreme phenomena in Kerr backgrounds.
The other applications may be null or unbound geodesics. 
We can apply our method to them with a few modifications. 
For null geodesics, we have to eliminate the mass term of the small body
 in the geodesics. 
For unbound geodesics, we can express the orbits in terms of the
 elliptic integrals using Mino time although there are no 
 fundamental frequencies for the orbits. 
However, it should be noted that we may have to improve computation time 
when we consider null or unbound geodesics 
because we have to trace the orbits for longer time than bound orbits cases. 
If we can not improve computation time, we may have to use 
both analytical solutions and numerical integration\cite{Rauch}. 
Finally, we note that we can not use the analytical solutions in this paper 
when the inclination angle from the equatorial plane of black hole is 
$\theta_{\rm inc}=\pi/2$. This is because $\Phi_\theta(\cos\theta)$ 
in (\ref{eq:geodesic_tau}) diverges when $\theta=0$, 
and its  elliptic integral  also diverges. 
We do not know how to  address this issue without any further approximation 
though one can solve it if one uses a post-Newtonian expansion~\cite{Ganz}. 
All of them will be discussed in a future work. 

\ack
We would like to thank Bala Iyer and Hiroyuki Nakano for useful comments. 
W.H. was supported by the JSPS Research Fellowships for Young Scientists, 
No.~1756 and also supported by the 21st Century COE program 
``Towards a New Basic Science; Depth and Synthesis'' 
at Osaka university from the Ministry of Education, Culture, Sports, 
Science and Technology of Japan. 
\appendix
\section{formulas of integrals of radial motion}
\label{sec:integral_formula}
In this appendix, we derive some formulas which are needed to obtain 
$\Upsilon_{t^{(r)}}$ and $\Upsilon_{\theta^{(r)}}$ in \sref{sec:Omega} 
and $t^{(r)}$ and $\phi^{(r)}$ in \sref{sec:ana_orbit}. 
In order to compute them, we have to investigate 
$\int^r_{r_2} \rmd r'/\{(r'-r_\pm)\sqrt{R(r')}\}$, 
$\int^r_{r_2} (r')^2\rmd r'/\sqrt{R(r')}$ and 
$\int^r_{r_2} \rmd r'/\{(r'-M)^2\sqrt{R(r')}\}$, see (\ref{eq:tr_phir_decom}).
Since $\int^r_{r_2} r'\rmd r'/\sqrt{R(r')}$ is derived in \sref{sec:Omega} and 
$\int^r_{r_2} \rmd r'/\{(r'-M)\sqrt{R(r')}\}$ can be derived from 
$\int^r_{r_2} \rmd r'/\{(r'-r_\pm)\sqrt{R(r')}\}$ when we set $r_\pm=M$, i.e. $a=M$, 
we do not show again these expressions  in this appendix.

As we derived $\int^r_{r_2} r'\rmd r'/\sqrt{R(r')}$ in \sref{sec:Omega}, 
it is useful to transform $r$ into $y_r$. Then we have the following relations.
\begin{eqnarray}
\eqalign{
\fl
\frac{1}{r-r_\pm}=
\frac{1}{r_3-r_\pm}\left[1-\frac{r_2-r_3}{r_2-r_\pm}\frac{1}{1-h_\pm y^2_r}\right],\\
\fl
r^2 = r^2_3 +\frac{(r_2-r_3)(r_2+3r_3)}{2}\frac{1}{1-h_r y^2_r}
+\frac{(r_2-r_3)^2}{4h_r}\left[\frac{1}{(y_r-h^{-1/2}_r)^2}+\frac{1}{(y_r+h^{-1/2}_r)^2}\right],\\
\fl
\left(\frac{r_3-M}{r-M}\right)^2 =
1 +\frac{1}{2}\frac{r_2-r_3}{r_2-M}\left(\frac{r_2-r_3}{r_2-M}-4\right)
\frac{1}{1-h_M y^2_r}\cr
+\frac{1}{4h_M}\left(\frac{r_2-r_3}{r_2-M}\right)^2
\left[\frac{1}{(y_r-h^{-1/2}_M)^2}+\frac{1}{(y_r+h^{-1/2}_M)^2}\right],
}
\end{eqnarray}
where $h_M=h_\pm(a=M)=(r_1-r_2)(r_3-M)/[(r_1-r_3)(r_2-M)]$.

Then it is straightforward to compute $\int^r_{r_2} \rmd r'/\{(r'-r_\pm)\sqrt{R(r')}\}$ as 
\begin{eqnarray}
\fl
\int^r_{r_2} \frac{\rmd r'}{(r'-r_\pm)\sqrt{R(r')}} =&
\frac{2}{(r_3-r_\pm)\sqrt{(1-{\C{E}}^2)(r_1-r_3)(r_2-r_4)}}\left[F(\arcsin y_r,k_r)\right.\cr
&-\left.\frac{r_2-r_3}{r_2-r_\pm}\Pi(\arcsin y_r,-h_\pm,k_r)\right]. 
\end{eqnarray}

However, we need a reformulation of the last terms of both 
$\int^r_{r_2} (r')^2\rmd r'/\sqrt{R(r')}$ and $\int^r_{r_2} \rmd r'/\{(r'-M)^2\sqrt{R(r')}\}$. 
If we set $J_n[c]=\int_0^y \rmd y'/\{(y'-c)^n\sqrt{\varphi(y')}\}$, 
where $\varphi(y)=(1-y^2)(1-k_r^2y^2)$, we can represent these terms as 
$J_2[c]+J_2[-c]$. Using reduction formula of the elliptic integrals, 
see section 17.1.5 in \cite{handbook}, 
we can express $J_2[c]+J_2[-c]$ in terms of the elliptic integrals as 
\begin{eqnarray}
\fl
J_2[c]+J_2[-c]=&
\frac{2}{\varphi(c)}\left\{\left[(2c^2-1)k_r^2-1\right]\Pi(\psi,-c^{-2},k_r)+(1-k_r^2c^2)F(\psi,k_r)-E(\psi,k_r)\right.\cr
&-\left.\left[\frac{y\sqrt{\varphi(y)}}{y^2-c^2}\right]_0^y\right\},
\end{eqnarray}
where $\psi=\arcsin y$.

Then we can express 
$\int^r_{r_2} (r')^2\rmd r'/\sqrt{R(r')}$ and $\int^r_{r_2} \rmd r'/\{(r'-M)^2\sqrt{R(r')}\}$ 
respectively as
\begin{eqnarray}
\fl
\int^r_{r_2} \frac{(r')^2}{\sqrt{R(r')}}\rmd r' =&
\frac{2}{\sqrt{(1-{\C{E}}^2)(r_1-r_3)(r_2-r_4)}}
\left[\frac{(r_3(r_1+r_2+r_3)-r_1r_2)}{2}F(\arcsin y_r,k_r)\right.\cr
&+\frac{(r_2-r_3)(r_1+r_2+r_3+r_4)}{2}\Pi(\arcsin y_r,-h_r,k_r)\cr
&+\frac{(r_1-r_3)(r_2-r_4)}{2}E(\arcsin y_r,k_r)\cr
&\left.+\frac{(r_1-r_3)(r_2-r_4)}{2}
\frac{y_r\sqrt{(1-y^2_r)(1-k_{r}^{2}y^2_r)}}{y^2_r-h_{r}^{-1}}
\right],
\end{eqnarray}
\begin{eqnarray}
\fl
\int^r_{r_2} \frac{(r_3-M)^2}{(r'-M)^2\sqrt{R(r')}}\rmd r' =&
\frac{2}{\sqrt{(1-{\C{E}}^2)(r_1-r_3)(r_2-r_4)}}\cr
&\times\left\{\left[1-\frac{1}{2}\frac{(r_1-r_3)(r_2-r_3)}{(r_1-M)(r_2-M)}\right]F(\arcsin y_r,k_r)\right.\cr
&+\frac{1}{2}\frac{r_2-r_3}{r_2-M}\left[\frac{r_1-r_3}{r_1-M}+\frac{r_2-r_3}{r_2-M}+\frac{r_4-r_3}{r_4-M}-4\right]\cr
&\times\Pi(\arcsin y_r,-h_{M},k_r)\cr
&+\frac{1}{2}\frac{(r_1-r_3)(r_2-r_4)(r_3-M)}{(r_1-M)(r_2-M)(r_4-M)}E(\arcsin y_r,k_r)\cr
&\left.+\frac{1}{2}\frac{(r_1-r_3)(r_2-r_4)(r_3-M)}{(r_1-M)(r_2-M)(r_4-M)}\frac{y_r\sqrt{(1-y^2_r)(1-k_{r}^{2}y^2_r)}}{y^2_r-h_{M}^{-1}}
\right\}.
\end{eqnarray}
\section{$|a|=M$ case}
\label{sec:a_M}
In this appendix, we show the analytical expressions of $\Gamma$, $\Upsilon_\phi$
$t^{(r)}$ and $\phi^{(r)}$ in the case $|a|=M$. 
Using the results of \ref{sec:integral_formula}, we derive them as 
\begin{eqnarray}
\Gamma=&
4M^2\C{E}+\frac{2a^2\C{E}z_{+}\Upsilon_{\theta}}{\pi\C{L}_{z}\sqrt{\epsilon_0 z_{+}}}
\left[K(k_\theta)-E(k_\theta)\right]\cr
&+\frac{2\Upsilon_{r}}{\pi\sqrt{(1-\C{E}^2)(r_1-r_3)(r_2-r_4)}}\left\{\frac{\C{E}}{2}\left[(r_3(r_1+r_2+r_3)-r_1r_2)K(k_r)\right.\right.\cr
&+(r_2-r_3)(r_1+r_2+r_3+r_4)\Pi(-h_r,k_r)\cr
&\left.+(r_1-r_3)(r_2-r_4)E(k_r)\right]+2M\C{E}\left[r_3K(k_r)+(r_2-r_3)\Pi(-h_r,k_r)\right]\cr
&+\frac{2M(4M^2\C{E}-a\C{L}_{z})}{r_3-M}
\left[K(k_r)-\frac{r_2-r_3}{r_2-M}\Pi(-h_{M},k_r)
\right]\cr
&+\frac{M^2(2M^2\C{E}-a\C{L}_{z})}{(r_3-M)^2}
\left[\left(2-\frac{(r_1-r_3)(r_2-r_3)}{(r_1-M)(r_2-M)}\right)K(k_r)\right.\cr
&+\frac{(r_1-r_3)(r_2-r_4)(r_3-M)}{(r_1-M)(r_2-M)(r_4-M)}E(k_r)\cr
&+\left.\left.\frac{r_2-r_3}{r_2-M}\left(\frac{r_1-r_3}{r_1-M}+\frac{r_2-r_3}{r_2-M}+\frac{r_4-r_3}{r_4-M}-4\right)\Pi(-h_{M},k_r)
\right]\right\},
\end{eqnarray}
\begin{eqnarray}
\Upsilon_{\phi} =& \frac{2\Upsilon_{\theta}}{\pi\sqrt{\epsilon_0 z_{+}}}
\Pi(-z_{-},k_\theta)+\frac{2a\Upsilon_{r}}{\pi\sqrt{(1-\C{E}^2)(r_1-r_3)(r_2-r_4)}}\cr
&\times\left\{\frac{2M\C{E}}{r_3-M}
\left[K(k_r)-\frac{r_2-r_3}{r_2-M}\Pi(-h_{M},k_r)
\right]\right.\cr
&+\frac{2M^2\C{E}-a\C{L}_{z}}{2(r_3-M)^2}
\left[\left(2-\frac{(r_1-r_3)(r_2-r_3)}{(r_1-M)(r_2-M)}\right)K(k_r)\right.\cr
&+\frac{(r_1-r_3)(r_2-r_4)(r_3-M)}{(r_1-M)(r_2-M)(r_4-M)}E(k_r)\cr
&+\frac{r_2-r_3}{r_2-M}\left(\frac{r_1-r_3}{r_1-M}+\frac{r_2-r_3}{r_2-M}+\frac{r_4-r_3}{r_4-M}-4\right)\cr
&\left.\left.\times\Pi(-h_{M},k_r)\right]\right\},
\end{eqnarray}
\begin{eqnarray}
\eqalign{
t^{(r)} =& \frac{2}{\sqrt{(1-\C{E}^2)(r_1-r_3)(r_2-r_4)}}\cr
&\times\left\{\frac{\C{E}}{2}\left[(r_2-r_3)(r_1+r_2+r_3+r_4)\tilde{\Pi}_r(\psi_r,-h_r,k_r)\right.\right.\cr
&+\left.(r_1-r_3)(r_2-r_4)\tilde{E}_r(\psi_r,h_r,k_r)\right]\cr
&+2M\C{E}(r_2-r_3)\tilde{\Pi}_r(\psi_r,-h_r,k_r)\cr
&-\frac{2M(4M^2\C{E}-a\C{L}_{z})}{r_3-M}
\frac{r_2-r_3}{r_2-M}\tilde{\Pi}_r(\psi_r,-h_{M},k_r)\cr
&+\frac{M^2(2M^2\C{E}-a\C{L}_{z})}{(r_3-M)^2}
\left[\frac{(r_1-r_3)(r_2-r_4)(r_3-M)}{(r_1-M)(r_2-M)(r_4-M)}\tilde{E}_r(\psi_r,h_{M},k_r)\right.\cr
&+\left.\left.\frac{r_2-r_3}{r_2-M}\left(\frac{r_1-r_3}{r_1-M}+\frac{r_2-r_3}{r_2-M}+\frac{r_4-r_3}{r_4-M}-4\right)\tilde{\Pi}_r(\psi_r,-h_{M},k_r)
\right]\right\},
}
\end{eqnarray}
\begin{eqnarray}
\eqalign{
\phi^{(r)} =& 
\frac{2a}{\sqrt{(1-\C{E}^2)(r_1-r_3)(r_2-r_4)}}
\left\{-\frac{2M\C{E}}{r_3-M}\frac{r_2-r_3}{r_2-M}\tilde{\Pi}_r(\psi_r,-h_{M},k_r)
\right.\cr
&+\frac{2M^2\C{E}-a\C{L}_{z}}{2(r_3-M)^2}
\left[
\frac{(r_1-r_3)(r_2-r_4)(r_3-M)}{(r_1-M)(r_2-M)(r_4-M)}\tilde{E}_r(\psi_r,h_{M},k_r)\right.\cr
&+\left.\left.\frac{r_2-r_3}{r_2-M}\left(\frac{r_1-r_3}{r_1-M}+\frac{r_2-r_3}{r_2-M}+\frac{r_4-r_3}{r_4-M}-4\right)\tilde{\Pi}_r(\psi_r,-h_{M},k_r)
\right]\right\}.
}
\end{eqnarray}


\section*{References}
\begin{thebibliography}{10}
\bibitem{Narayan}
Narayan R 2005 {\it New J. Phys.} {\bf 7} 199 

\bibitem{Chandra}
Chandrasekhar S 1983 {\it Mathematical theory of black holes} (Oxford: Oxford University  Press)

\bibitem{wilkins} Wilkins D C 1972 {\it Phys.\ Rev.\ \rm D} {\bf 5} 814 

\bibitem{Hughes2} 
Hughes S A 2001 {\it Phys.\ Rev.\ \rm D} \textbf{63} 064016

\bibitem{Barausse}
Barausse E, Hughes S A and Rezzolla L 2007 {\it Phys. Rev. \rm D} \textbf{76} 044007 

\bibitem{GK} 
Glampedakis K and Kennefick D 2002 {\it Phys. Rev. \rm D} \textbf{66} 044002 

\bibitem{Celestial}
Schmidt W 2002 {\it Class. Quantum Grav.} \textbf{19} 2743

\bibitem{Mino:2003yg}
Mino Y 2003 {\it Phys. Rev. \rm D} \textbf{67} 084027 

\bibitem{Drasco:2004}
Drasco S and Hughes S A 2004 {\it Phys. Rev. \rm D} \textbf{69} 044015

\bibitem{Kostas_Review}
Glampedakis K 2005 {\it Class. Quantum Grav.} \textbf{22} S605

\bibitem{LISA} {\it LISA web page} http://lisa.jpl.nasa.gov/

\bibitem{Rauch} 
Rauch K P and Blandford R D 1994 {\it Astrophys. J.} \textbf{421} 46

\bibitem{Dexter} 
Dexter J and Agol E 2009 {\it Astrophys. J.} \textbf{696} 1616

\bibitem{handbook} 
Abramowitz M and Stegun I A (eds) 1972 \textit{Handbook of Mathematical Functions} (New York: Dover)

\bibitem{carlson1988}
Carlson B C 1988 {\it Math. Comp.} \textbf{51} 267

\bibitem{Bardeen} 
Bardeen J M 1973 \textit{Black Holes} ed C DeWitt and B S DeWitt (New York: Gordon and Breach Science Publishers)

\bibitem{Recipes} 
Press W H, Teukolsky S A, Vetterling W T and Flannery B P 1992 \textit{Numerical Recipes in C} (Cambridge: Cambridge University Press)

\bibitem{SY2011} 
C. F. Sopuerta and N. Yunes 2011 {\it Phys. Rev. \rm D} \textbf{84} 124060

\bibitem{BPT} 
Bardeen J M, Press W H and Teukolsky S A 1972 {\it Astrophys. J.} \textbf{178} 347 

\bibitem{CKP} 
Cutler C, Kennefick D~and Poisson E 1994 {\it Phys. Rev. \rm D} \textbf{50} 3816 

\bibitem{Hughes1}
Hughes S A 2000 {\it Phys. Rev. \rm D} \textbf{61} 084004\\
Hughes S A 2001 {\it Phys.\ Rev.\ \rm D} \textbf{63} 049902(E)\\
Hughes S A 2002 {\it Phys.\ Rev.\ \rm D} \textbf{65} 069902(E)\\
Hughes S A 2003 {\it Phys.\ Rev.\ \rm D} \textbf{67} 089901(E) 

\bibitem{Ganz}
Ganz K, Hikida W, Nakano H, Sago N and Tanaka T 2007 {\it Prog. Theor. Phys.} \textbf{117} 1041

\bibitem{Kraniotis:2007}
Kraniotis G V 2007 {\it Class. Quantum Grav.} \textbf{24} 1775

\bibitem{carlson1970}
Zill D G and Carlson B C 1970 {\it Math. Comp.} \textbf{24} 199

\bibitem{Teukolsky:1973ha}
Teukolsky S A 1973 {\it Astrophys. J.} \textbf{185} 635

\bibitem{chapter}
Mino Y, Sasaki M, Shibata M, Tagoshi H and Tanaka T 1997 {\it Prog. Theor. Phys. Suppl.} \textbf{118} 1

\bibitem{ST} 
Sasaki M and Tagoshi H 2003 {\it Living Rev. Relativity} \textbf{6} 6

\bibitem{Drasco:2006}
Drasco S and Hughes S A 2006 {\it Phys. Rev. \rm D} \textbf{73} 024027 

\bibitem{FHT}
Fujita R, Hikida W and Tagoshi H 2009 {\it Prog. Theor. Phys.} \textbf{121} 843

\bibitem{Hughes:2005qb}
Hughes S A Drasco S Flanagan E E and Franklin J 2005 {\it Phys. Rev. Lett.} \textbf{94} 221101

\bibitem{Levin1}
Levin J and Perez-Giz G 2008 {\it Phys. Rev. \rm D} \textbf{77} 103005 

\bibitem{Levin2}
Levin J and Grossman B 2009 {\it Phys. Rev. \rm D} \textbf{79} 043016

\end{thebibliography}
\end{document}